\documentclass[sigconf]{acmart} 

\usepackage{amsmath}
\usepackage{multirow}
\usepackage{float}
\usepackage{tikz}
\definecolor{OliveGreen}{rgb}{0.22, 0.6, 0.21}

\usepackage{placeins}
\usepackage{afterpage}
\usepackage{multicol}
\AtBeginDocument{%
  \providecommand\BibTeX{{%
    \normalfont B\kern-0.5em{\scshape i\kern-0.25em b}\kern-0.8em\TeX}}}

\copyrightyear{2022}
\acmYear{2022}
\setcopyright{acmlicensed}\acmConference[SIGGRAPH '22 Conference Proceedings]{Special Interest Group on Computer Graphics and Interactive Techniques Conference Proceedings}{August 7--11, 2022}{Vancouver, BC, Canada}
\acmBooktitle{Special Interest Group on Computer Graphics and Interactive Techniques Conference Proceedings (SIGGRAPH '22 Conference Proceedings), August 7--11, 2022, Vancouver, BC, Canada}
\acmPrice{15.00}
\acmDOI{10.1145/3528233.3530700}
\acmISBN{978-1-4503-9337-9/22/08}

\acmConference[SIGGRAPH '22 Conference Proceedings]{SIGGRAPH '22 Conference Proceedings}{August 07--11,
  2022}{Vancouver, BC, Canada}
\acmSubmissionID{papers\_195}

\begin{document}
\title{Neural Shadow Mapping}


\author{Sayantan Datta}
\email{sayantan.datta@mail.mcgill.ca}
\affiliation{%
  \institution{McGill University}
  \city{Montreal}
  \state{QC}
  \country{Canada}
}
\affiliation{%
  \institution{Meta Reality Labs}
  \city{Redmond}
  \state{WA}
  \country{USA}
}

\author{Derek Nowrouzezahrai}
\email{derek@cim.mcgill.ca}
\affiliation{%
  \institution{McGill University}
  \city{Montreal}
  \state{QC}
  \country{Canada}
}

\author{Christoph Schied}
\email{cschied@nvidia.com}
\author{Zhao Dong}
\email{zhaodong@fb.com}
\affiliation{%
  \institution{Meta Reality Labs}
  \city{Redmond}
  \state{WA}
  \country{USA}
}



\renewcommand{\shortauthors}{Datta, et al.}

\begin{abstract}
  We present a neural extension of basic shadow mapping for fast, high quality hard and soft shadows. We compare favorably to fast pre-filtering shadow mapping, all while producing visual results on par with ray traced hard and soft shadows. We show that combining memory bandwidth-aware architecture specialization and careful temporal-window training leads to a fast, compact and easy-to-train neural shadowing method. Our technique is memory bandwidth conscious, eliminates the need for post-process temporal anti-aliasing or denoising, and supports scenes with dynamic view, emitters and geometry while remaining robust to unseen objects.
\end{abstract}



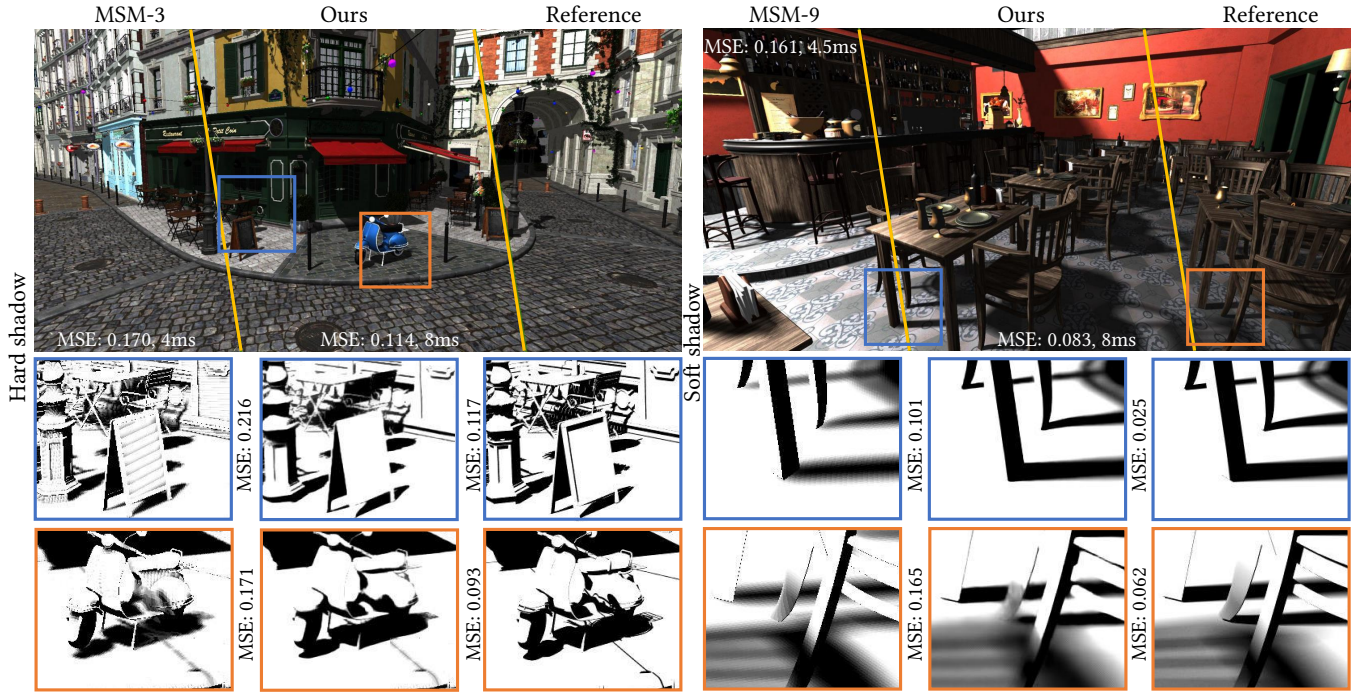
\begin{teaserfigure}
\begin{tikzpicture}
    \node[anchor=south west,inner sep=0] at (0,0){\includegraphics[width=\textwidth, trim={0cm 1.15cm 0cm 1.0cm},clip]{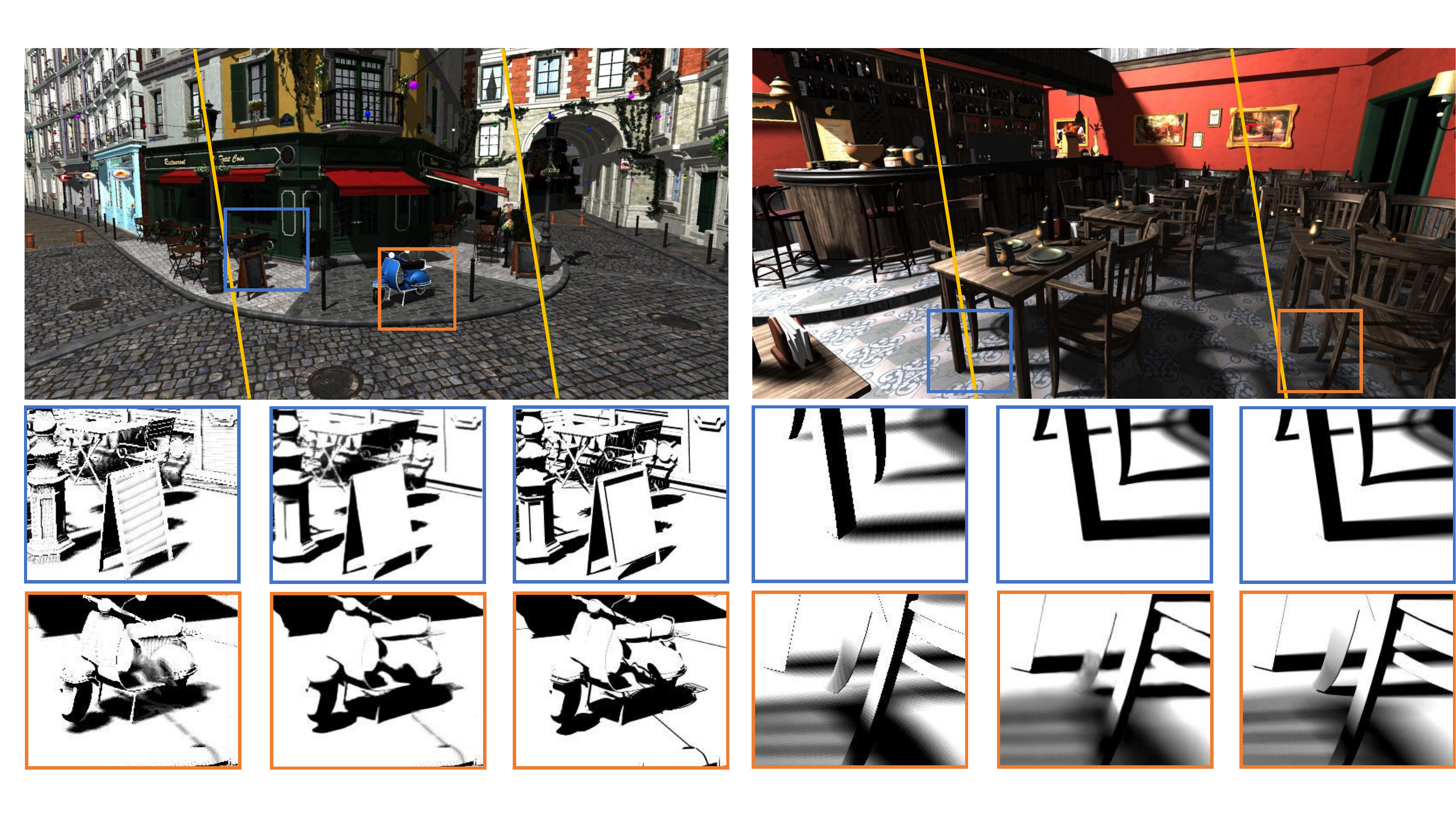}}; 
    \node[anchor=south west] at (1,9.45 - 0.65) { \textcolor{black}{MSM-3} };
    \node[anchor=south west] at (4,9.45 - 0.65) { \textcolor{black}{Ours} };
    \node[anchor=south west] at (7,9.45 - 0.65) { \textcolor{black}{Reference} };
    \node[anchor=south west] at (9.7,9.45 - 0.65) { \textcolor{black}{MSM-9} };
    \node[anchor=south west] at (13,9.45 - 0.65) { \textcolor{black}{Ours} };
    \node[anchor=south west] at (16,9.45 - 0.65) { \textcolor{black}{Reference} };
    \node[anchor=south west, rotate=90] at (0.3,4.4 - 0.65) { \textcolor{black}{Hard shadow} };
    \node[anchor=south west, rotate=90] at (9.32 - 0.05,4.4 - 0.65) { \textcolor{black}{Soft shadow} };
    \node[anchor=south west, rotate=90] at (3.3,3.2 - 0.65) { \textcolor{black}{\small MSE: 0.216} };
    \node[anchor=south west, rotate=90] at (3.33,0.95 - 0.65) { \textcolor{black}{\small MSE: 0.171} };
    \node[anchor=south west, rotate=90] at (6.35,3.2 - 0.65) { \textcolor{black}{\small MSE: 0.117} };
    \node[anchor=south west, rotate=90] at (6.35,0.95 - 0.65) { \textcolor{black}{\small MSE: 0.093} };
    \node[anchor=south west, rotate=90] at (12.32 - 0.1,3.2 - 0.65) { \textcolor{black}{\small MSE: 0.101} };
    \node[anchor=south west, rotate=90] at (12.32 - 0.1,0.95 - 0.65) { \textcolor{black}{\small MSE: 0.165} };
    \node[anchor=south west, rotate=90] at (15.3 - 0.1,3.2 - 0.65) { \textcolor{black}{\small MSE: 0.025} };
    \node[anchor=south west, rotate=90] at (15.3 - 0.1,0.95 - 0.65) { \textcolor{black}{\small MSE: 0.062} };
    \node[anchor=south west] at (0.5,5.1 - 0.65) { \textcolor{white}{\small MSE: 0.170, 4ms} };
    \node[anchor=south west] at (4,5.1 - 0.65) { \textcolor{white}{\small MSE: 0.114, 8ms} };
    \node[anchor=south west] at (9.1,9.0 - 0.65) { \textcolor{white}{\small MSE: 0.161, 4.5ms} };
    \node[anchor=south west] at (13,5.1 - 0.65) { \textcolor{white}{\small MSE: 0.083, 8ms} };
\end{tikzpicture}
\caption{Our hard and soft shadowing method approaches the quality of offline ray tracing whilst striking a favorable position on the performance-accuracy spectrum. On the high-performance end, we produce higher quality results than $n \times n$ Moment Shadow Maps (MSM-$n$). We require only vanilla shadow mapping inputs to generate visual (and temporal) results that approach ray-traced reference, surpassing more costly denoised interactive ray-traced methods.}
\label{fig:teaser}
\end{teaserfigure}
\vspace{-10pt}
\maketitle

\section{Introduction}

Shadows provide important geometric, depth and shading cues. Real-time hard and soft shadow rendering remains a challenge, especially on resource-limited systems. Pre-filtering based methods \cite{MSM15, VSM06, ConvSM07} are fast but approximate. They are prone to light leaking artifacts, reduced shadow contrast, and limited contact hardening. Interactive ray-tracing \cite{HWRT09} coupled with post-process denoising \cite{SVGF17, ChaitanyaDenoise17} and upscaling \cite{NeuralSS} can deliver high quality dynamic shadows, but even the fastest GPU ray-tracers fall short of the performance demands of interactive graphics. Low ray-tracing hardware adoption and the added engineering complexity of integrating GPU ray tracers into rasterization-based pipelines is another limitation. Pre-computation based methods \cite{PRT02, NeRF20} do not generally support dynamic objects or near-field light transport, and require significant memory.

We propose a machine learning-based method that generates high quality hard and soft shadows for dynamic objects in real-time. Our approach does not require ray-tracing hardware, has high performance (< 6ms), requires little memory (< 1.5MBs), and is easy to deploy on commodity low-end GPU hardware. We use the output of ``vanilla'' rasterization-based shadow mapping (i.e., no cascades, etc.) to hallucinate temporally-stable hard and soft shadows. We design a compact neural architecture based on the statistics of penumbra sizes in a diversity of scenes. The network admits rapid training and generalizes to unseen dynamic objects. We demonstrate improved quality over state of the art in high-performance pre-filtering based methods while retaining support for dynamic scenes and approaching reference-quality results.

We show that careful feature engineering, application and memory aware architecture design, combined with a novel temporal stability loss results in a system with many favorable properties: apart from compactness and high-performance, our output precludes the need for post-process temporal anti-aliasing (TAA), further reducing the renderer's bandwidth requirements. We parameterize our network by emitter size, allowing us to encode both hard and soft shadow variation into a single trained net. We demonstrate its effectiveness on several scenes with dynamic geometry, camera, and emitters. Our results are consistently better than workhorse interactive methods, and they also rival much slower (and more demanding, system- and hardware-wise) interactive ray-tracing and denoising-based pipelines. Finally, we discuss scene dependent optimizations that further reduce our network size and runtime.

\section{Related work}
Shadow mapping \cite{SM1978} and its variants are efficient shadowing methods for point and directional lights in dynamic scenes. Shadow map resolution and projection leads to shadow aliasing artifacts, with solutions (e.g., depth biasing~\cite{ShadowSlopeBias04,AdaptiveDepthBias14}) leading to secondary issues and trade-offs. Modern shadow mapping relies on delicately engineered systems that combine many \textit{cascaded maps}~\cite{PSSM06, CSM07}. Here, we refer readers to a comprehensive survey~\cite{ShadowBook}.

Filtering-based methods prefilter (in emitter-space) depth-based visibility to reduce aliasing. One simple such method weights nearby depth samples~\cite{PCF87}; this \textit{percentage closer filtering} remains a commonly used technique in interactive applications, with a recent variant that modulates the filter size based on the relative blocker and receiver positions is used to approximate soft shadows~\cite{PCSS05}. More recently, a new class of filtering methods replace binary depth samples with statistical proxies, allowing for more sophisticated pre-filtering~\cite{ConvSM07,VSM06, ESM08} and coarse approximations of soft shadows. Moment shadow maps \cite{MSM15} are the state of the art of these methods, but it can suffer from banding, aliasing, light leaking in scenes with high depth complexity.

Screen-space methods treat G-buffers, including screen-projected shadow map data, leveraging image-space locality and GPU parallelization for efficient filtering in a deferred shading pipeline. Here, accurate filtering here requires the determination of an potentially-anisotropic filter kernel (due to perspective distortion), and so depends non-linearly on the viewing angle \cite{ScreenSpaceAniSS} and pixel depths \cite{SSPCSS10}. Our method similarly treats image-space G-buffer data, but we instead \textit{learn} compositional filters from data. High-fidelity soft shadows also benefit from occluder depth estimates from both the emitter and shade point, of which only the former is readily available from the shadow map and the latter can be approximated using min- \shortcite{SSSSGpuPro360} or average-filtering \shortcite{SSPCSS10} of the projected shadow map. Again, we rely on learning compositions of convolution and pooling layers to model (the effects of) these depth estimates.

Ray tracing hardware opens up an exciting new avenue for dynamic hard and soft shadows. These methods, however, remain power-inefficient and typically require post-process denoising (traditional \cite{SVGF17} or machine learning-based \cite{NeuralLayerDenoising20, ChaitanyaDenoise17}) and TAA~\cite{NeuralSS, Dlss19} to attain modest interactivity.

\section{Overview}

\begin{figure}[b!]
\begin{tikzpicture}
    \node[anchor=south west,inner sep=0] at (0,0) {
    \includegraphics[width=9.0cm, trim={0cm 5.95cm 0 0.3cm},clip]{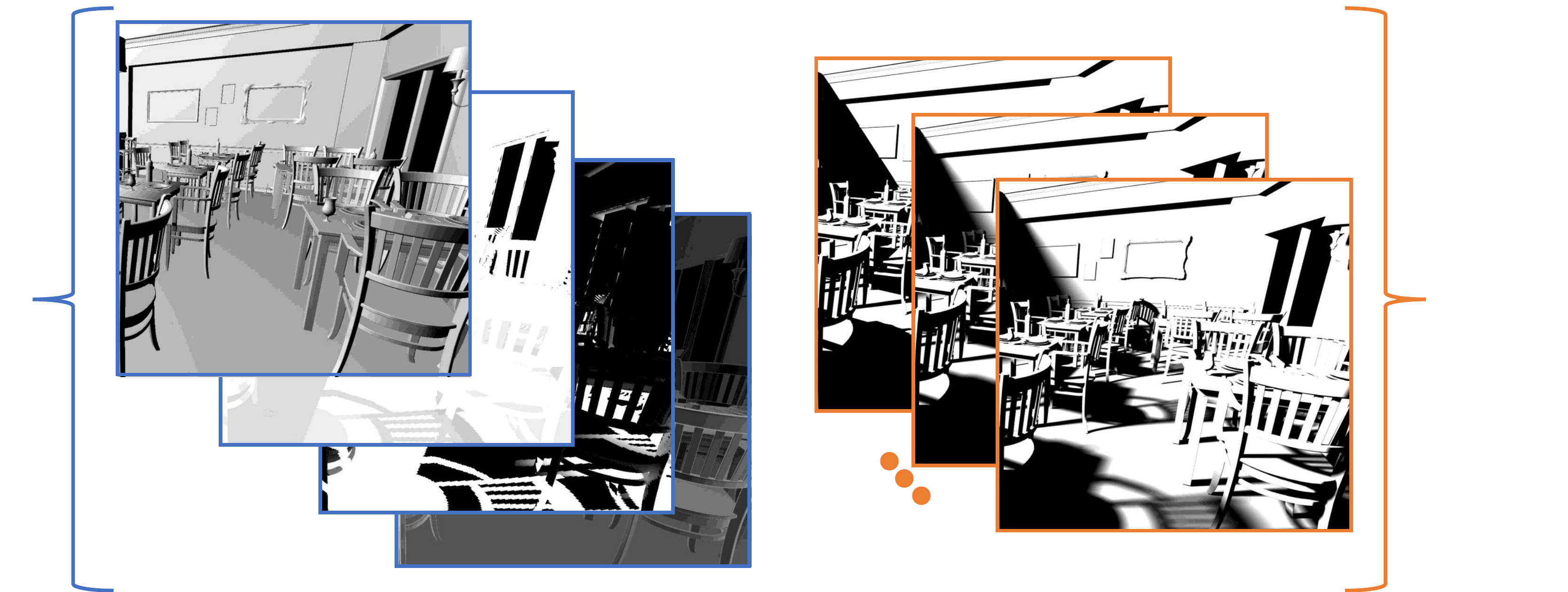}}; 
    \node[anchor=south west, rotate=90] at (0.25, 2.1 - 1.55) { \textcolor{black}{\small Rasterized features} };
    \node[anchor=south west, rotate=90] at (8.55, 2.1 - 1.55) { \textcolor{black}{\small Ray-traced targets} };
    \node[anchor=south west] at (1.0, 4.78-1.55) { \textcolor{black}{$c_e + r_e$} };
    \node[anchor=south west] at (2.63, 4.4-1.55) { \textcolor{black}{$z/z_f$} };
    \node[anchor=south west] at (3.23, 4.0-1.55) { \textcolor{black}{$z - z_f$} };
    \node[anchor=south west] at (3.83, 3.65-1.55) { \textcolor{black}{$c_c/d$} };
    \node[anchor=south west] at (4.73, 4.58 - 1.55) { \textcolor{black}{Hard shadow} };
    \node[anchor=south west] at (5.73, 1.58 - 1.55) { \textcolor{black}{Soft shadow} };
    
\end{tikzpicture}
\caption{Visualizing supervised learning pairs. The network inputs are the rasterization buffers modulated by the size of emitter ($r_e$). The targets are generated using ray-tracing according to the corresponding emitter size.}
\label{fig:supervisedLearningPairs}
\end{figure}
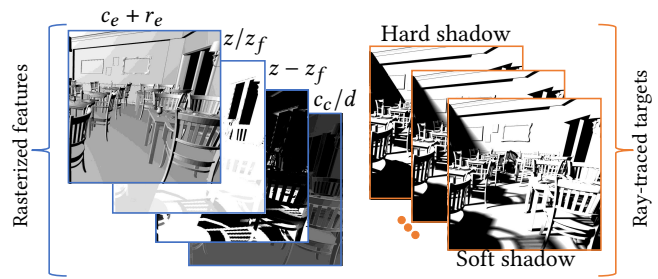

Overall, our approach is straightforward; we generate a set of screen-space buffers using a G-buffer and a shadow mapping pass before passing them as inputs to our network. The output of the network is compared against ray-traced shadows as target during training. Although straightforward, simply using a UNet \cite{UNet15, DeepShading16} without our proposed training and optimization methodology yields a network that is temporally unstable, bandwidth limited, heavy (>25MBs) and too slow (>100ms) for real-time use. As such, our methodology is focused on making conscious choices to preserve memory bandwidth while having minimal impact on quality.

We train our network using screen-space buffers as features and corresponding ray-traced shadows as targets. The approach allows for easy integration into the rendering pipeline while providing room for integration (possible future work) into supersampling \cite{NeuralSS} and neural-shading \cite{DeepShading16}. Our approach is also suitable for tiled rendering - popular among mobile devices. We develop a methodology to select a compact set of features that preserve necessary information without increasing memory bandwidth. We then show a simple technique to encode shadows with variable emitter size in a single network. We design a loss function to enhance the temporal stability of network without using historical buffers, thus further reducing bandwidth requirements. We provide several network architecture optimizations aimed at reducing memory and compute requirements. While our general architecture supports flexible emitter sizes, we show a recipe to further optimize our network for a fixed emitter size, enabling further trimming of network layers.

\begin{figure}[t!]
 \begin{tikzpicture}
    \node[anchor=south west,inner sep=0] at (0.8,0) {
    \includegraphics[width=11.5cm, trim={0cm 8.75cm 0cm 0.7cm},clip]{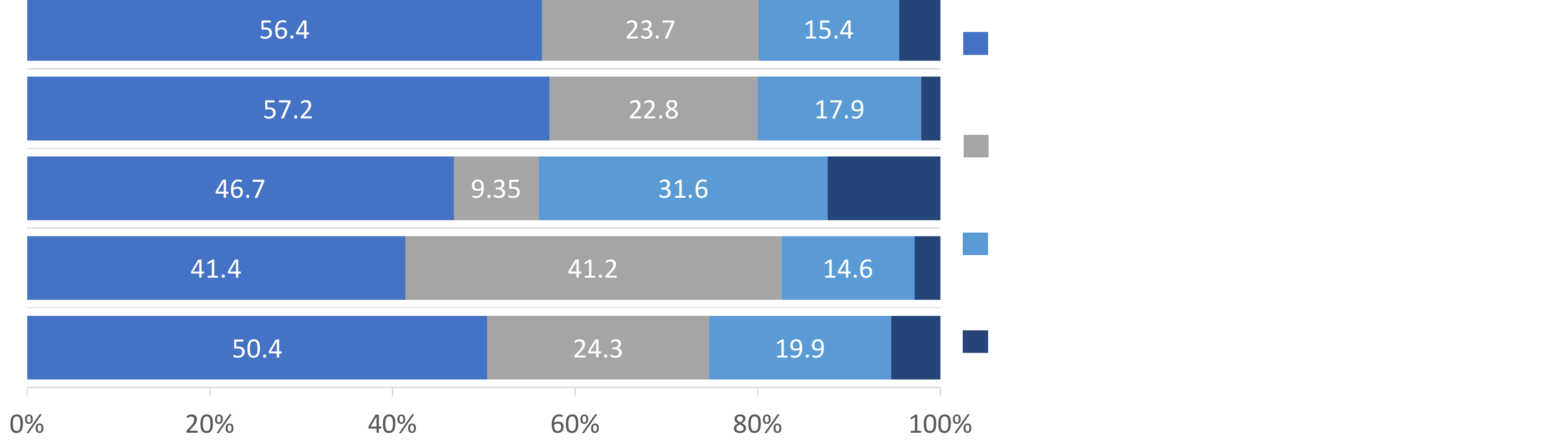}};
    \node[anchor=south west] at (0.0, 2.9 - 0.1) { \textcolor{black}{\small Bistro} };
      \node[anchor=south west] at (0.1, 2.35 - 0.1) { \textcolor{black}{\small Conf.} };
      \node[anchor=south west] at (0.0, 1.75 - 0.1) { \textcolor{black}{\small Sponza} };
      \node[anchor=south west] at (0.0, 1.225) { \textcolor{black}{\small Living} };
      \node[anchor=south west] at (0.0, 1.0) { \textcolor{black}{\small Room} };
      \node[anchor=south west] at (0.2, 0.45) { \textcolor{black}{\small Avg.} };
      
      \node[anchor=south west, rotate=90] at (8.5, 2.65 - 0.1) { \textcolor{black}{\small$z-z_f$} };
       \node[anchor=south west, rotate=90] at (8.5, 1.9 - 0.1) { \textcolor{black}{\small$z/z_f$} };
       \node[anchor=south west, rotate=90] at (8.5, 1.35 - 0.1) { \textcolor{black}{\small$c_e$} };
       \node[anchor=south west, rotate=90] at (8.5, 0.4 - 0.1) { \textcolor{black}{\small$c_c/d$} };
\end{tikzpicture}
\caption{\label{fig:relativeSensitivity} Relative sensitivity of the selected features for various scenes.}
\end{figure}
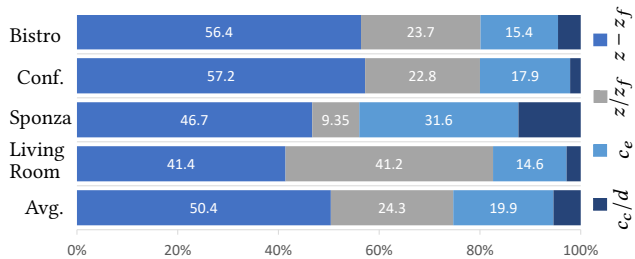
\subsection{Supervised training pairs}

We use supervised learning to train our neural network. The training examples are generated using rasterization and ray-tracing for features and targets respectively. The rasterization pipeline includes a G-buffer pass followed by a shadow mapping pass. Together they generate the following screen-space buffers:
\begin{itemize}
    \item view-space depth $d$ and normal $\mathbf{n}$,
    \item emitter-to-occluder depth $z$ and emitter-space normal $\mathbf{n_e}$,
    \item pixel-to-emitter distance $z_f$, the emitter radius (size) $r_e$ for spherical sources, and
    \item dot products $\{c_e, c_c\}$ of $\mathbf{n}$ with the emitter direction and $\mathbf{n}$ with the viewing direction.
\end{itemize}
The ray-tracing pass generates converged images of hard and soft shadows using a brute force Monte-Carlo sampling and a mild Gaussian filter. An 8x multi-sample anti-alising (MSAA) is also applied to the ray-traced targets. We do not however use MSAA in the rasterization pipeline as we expect the network to implicitly learn anti-aliasing from the target images.

\subsubsection{Softness control} We train a single network to predict a range of shadows with varying softness. Note that the same input from the rasterization is used to generate both hard and soft shadows. The softness is controlled (on a continuous scale) using a scalar parameter indicating the size of the emitter. For training, the emitter sizes are encoded as integer textures between 0 to 4, where 0 indicates a point light and 4 indicates the largest emitter size (diameter 50 cm). Rather than passing the scalar values to the network as an additional constant screen-space buffer, a more bandwidth efficient approach is to add the scalar as a dc-offset to an already existing buffer. We choose the cosine texture ($c_e$) to add the emitter-size ($r_e$) dc-offset to. The network targets are also changed corresponding to the selected emitter size. See figure~\ref{fig:supervisedLearningPairs}. During inference, the network accepts a scalar between 0 and 4 and intrinsically interpolates across discrete emitter values the network is trained with.

\subsection{Feature selection}

While dumping the content of the rasterization pass through a network works, it is bandwidth inefficient and adds a 2.5ms penalty to the cost of evaluating the network. The inefficient technique involves adding a feature extraction network, cascaded before the main network and training the two network end to end. The feature extraction network consisting of several layers of 1x1 convolutions, compresses all 15 channels of rasterizer output down to 4. A 1x1 convolution layer acts as fully connected layer across channels without performing any convolution across pixels. We tested a 2-layer deep 1x1 convolution network, recorded an overall error of $6.64\times10^{-3}$ across a suite of scenes involving hard and soft shadows.

Our approach eliminates the need for a feature extraction network by systematic evaluation and selection of the rasterization output buffers. We first introduce the notion of \textit{sensitivity}, a metric we use to quantify the importance of a feature. \textit{Sensitivity} measures a change in the network output due to a small perturbation in the input. Intuitively, sensitivity is lower if a channel's contribution in explaining the output variation is lower. Absolute sensitivity $S_i$ for the $i^{th}$ input channel $f_i$ is given by 
\begin{equation}
    S_i = \mathbb{E}\left[\frac{\left(\phi(f_i + \epsilon_i) - \phi(f_i)\right)}{0.1\sigma_i}\right], \;\;
    \epsilon_i \sim \mathcal{N}(0, 0.1\sigma_i)
\end{equation}
where $\phi$ is the network, the random perturbation texture $\epsilon_i$ is obtained by sampling a normal distribution. The standard deviation $\sigma_i$ corresponding to the $i^{th}$ channel is empirically estimated by aggregating all pixels in the dataset for that channel. The formula is repeated several times to reduce sampling noise. To compare the sensitivities across different training instances, we compute \textbf{relative sensitivity} as $s_i = S_i / \sum_i  S_i$.

\begin{figure}[t!]
\begin{tikzpicture}
    \node[anchor=south west,inner sep=0] at (0,0) {
    \includegraphics[width=9.5cm, trim={0cm 3.9cm 0cm 1.1cm},clip]{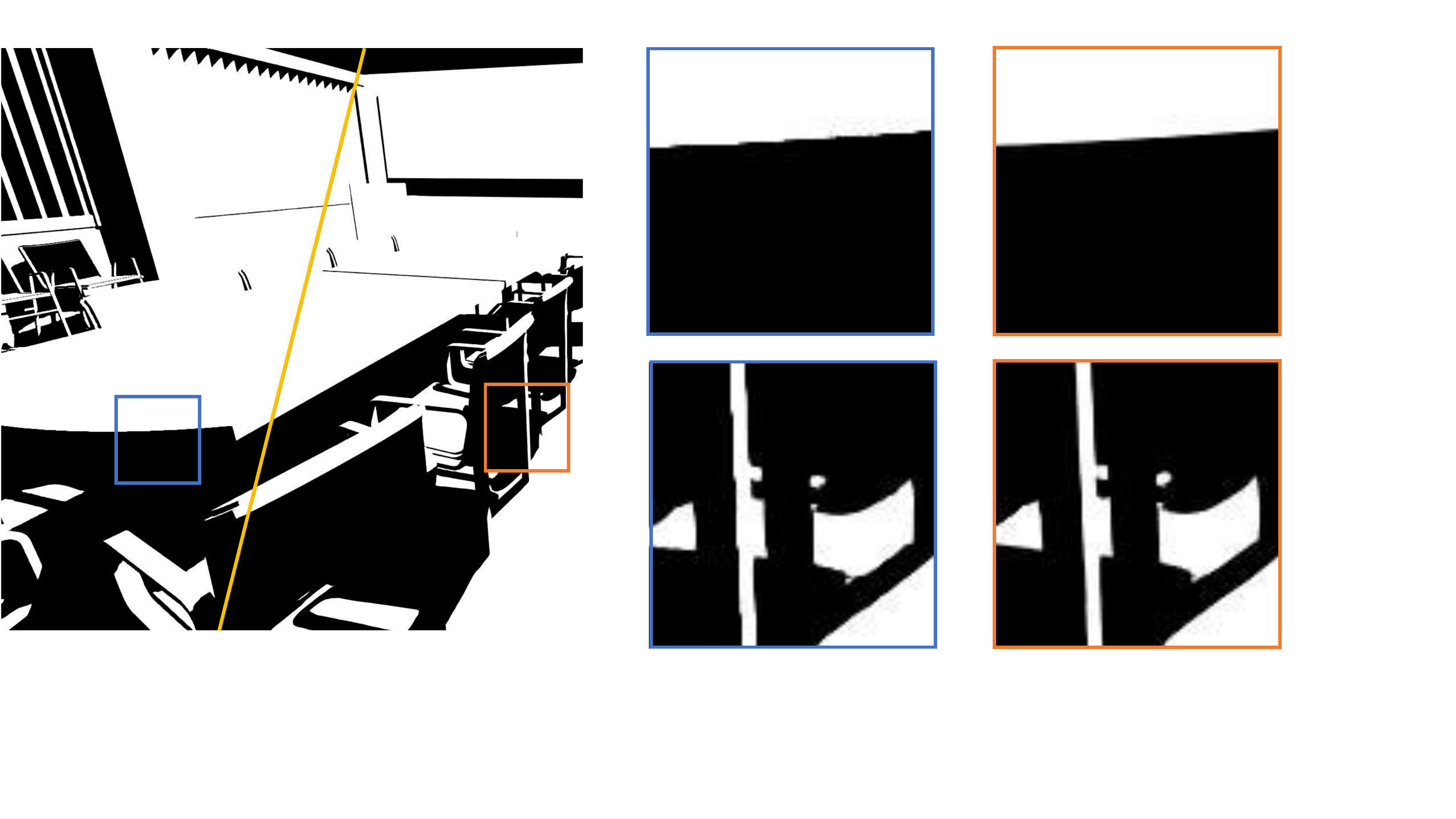}};
    \node[anchor=south west] at (0.5, 4.0-0.15) { \textcolor{black}{Vgg-Off} };
    \node[anchor=south west] at (2.5, 4.0-0.15) { \textcolor{black}{Vgg-On} };
    \node[anchor=south west] at (4.5, 4.0-0.15) { \textcolor{black}{Vgg-Off} };
    \node[anchor=south west] at (6.8, 4.0-0.15) { \textcolor{black}{Vgg-On} };
    \node[anchor=south west] at (0.67, 3.4-0.15) { \textcolor{black}{MSE: 0.038} };
    \node[anchor=south west] at (2.3, 3.4-0.15) { \textcolor{black}{MSE: 0.035} };
    \node[anchor=south west, rotate=90] at (4.25, 2.35-0.15) { \textcolor{black}{MSE: 0.020} };
    \node[anchor=south west, rotate=90] at (4.25, 0.3-0.15) { \textcolor{black}{MSE: 0.057} };
    \node[anchor=south west, rotate=90] at (6.5, 2.35-0.15) { \textcolor{black}{MSE: 0.007} };
    \node[anchor=south west, rotate=90] at (6.5, 0.3-0.15) { \textcolor{black}{MSE: 0.046} };
\end{tikzpicture}
\caption{Effect of VGG loss on the final output. VGG loss produces sharper edges for geometry and shadow silhouettes.}
\label{fig:effectOfVgg}
\end{figure}
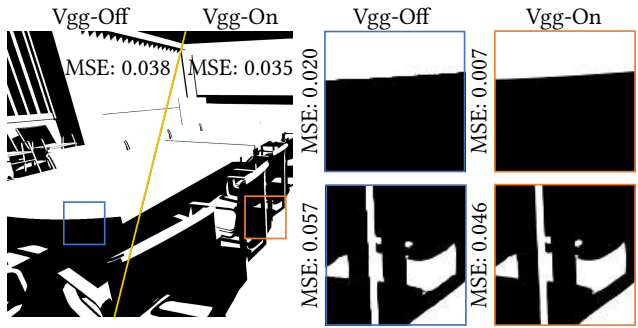

Armed with relative sensitivity as our yardstick, our problem is thus selecting a subset of features from a set of features $U = \left\{d, \mathbf{n}, z, \mathbf{n_e}, z_f, c_e, c_c \right\} + \left\{z - z_f, z / z_f, c_c / d, \mathbf{n} \cdot \mathbf{n_e}\right\}$. The first set of buffers are obtained directly from the rasterization while the second set is a composition from the first set. We take a tiered approach for selecting the best features. In the first pass, we train our network with all buffers in set $U$ and reject buffers with low relative sensitivity. We repeat the process until all buffers have sensitivity higher than 1.5\%. Refer to supplemental\footnote{Attached herewith at the end} material, section 1.0.2 for more details. Our final set of buffers is as shown in figure \ref{fig:relativeSensitivity}. We obtain nearly the same error ($6.67\times10^{-3}$) as having a feature extraction while saving an extra 2.5ms. 

\subsection{Loss function and temporal stability}

The loss function plays two main role in defining the characteristics of our network. It shapes the network to better fit the hard edges for shadow silhouette and geometry, essentially performing post-process anti-aliasing. Second, our loss function improves temporal stability without using any historical buffers for training and inference. Our approach not only saves memory bandwidth but also enables easier integration into tiled renderers.

We achieve the first objective using a weighted combination of per-pixel difference and VGG-19 \shortcite{Vgg19} perceptual loss. The effect of VGG loss on the final output is shown in figure \ref{fig:effectOfVgg} and clearly shows the anti-aliasing effect of VGG-19 on hard edges.

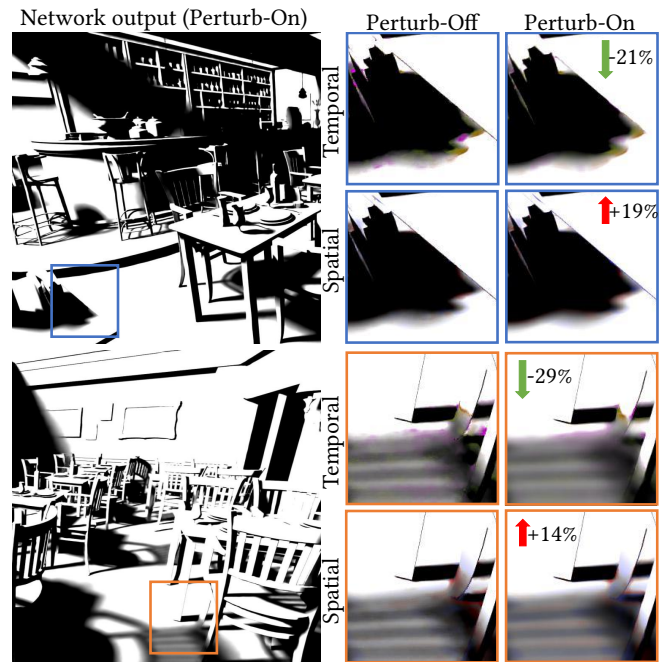
\begin{figure}[t!]
\begin{tikzpicture}
    \node[anchor=south west,inner sep=0] at (0,0) {
\includegraphics[width=14.9cm, trim={0cm 0.0cm 0cm 0.0cm},clip]{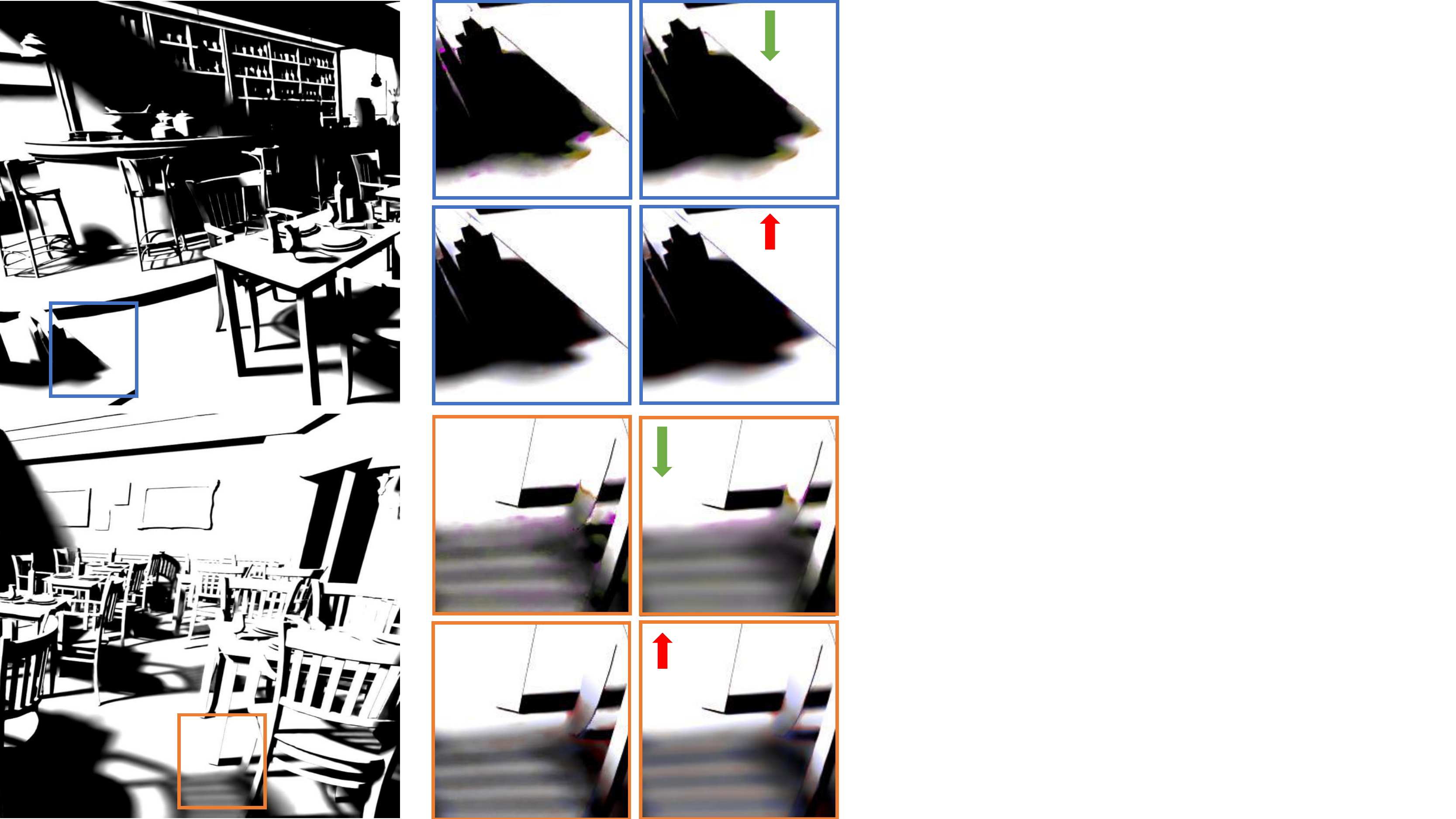}}; 
    \node[anchor=south west] at (0, 8.3) { \textcolor{black}{Network output (Perturb-On)} };
   \node[anchor=south west] at (4.6, 8.3) { \textcolor{black}{Perturb-Off} };
   \node[anchor=south west] at (6.7, 8.3) { \textcolor{black}{Perturb-On} };
   \node[anchor=south west, rotate=90] at (4.525, 6.6) { \textcolor{black}{Temporal} };
   \node[anchor=south west] at (7.81, 7.85) { \textcolor{black}{-21\%} };
   \node[anchor=south west, rotate=90] at (4.525, 4.8) { \textcolor{black}{Spatial} };
   \node[anchor=south west] at (7.81, 5.8) { \textcolor{black}{+19\%} };
   \node[anchor=south west, rotate=90] at (4.525, 2.2) { \textcolor{black}{Temporal} };
   \node[anchor=south west] at (6.73, 3.6) { \textcolor{black}{-29\%} };
   \node[anchor=south west, rotate=90] at (4.525, 0.4) { \textcolor{black}{Spatial} };
   \node[anchor=south west] at (6.73, 1.5) { \textcolor{black}{+14\%} };
\end{tikzpicture}
\caption{Comparing the temporal and spatial effect of perturbation loss. The application of perturbation loss reduces temporal instability while causing an increase in spatial blur as shown in the cutouts. We measure temporal instability by comparing the network output between consecutive frames while we measure spatial error by comparing the network output with reference. Temporal instability and spatial errors are represented using false colors purple/gold and red/blue respectively.}
\label{fig:effectOfPerturbationLoss}
\end{figure}

Existing methods typically improves temporal stability using historical buffers to better support the network during inference \cite{NeuralSS, Dlss19} and also to reshape the loss function \cite{DeepFovea, DeepCloth19} during training. In our case, we do not use historical buffers but use random perturbations of the input buffers for reshaping the loss landscape during training. Temporal instabilities arise due to shadow-map aliasing, where shadow map texels do not align one-to-one with screen pixels. As such small, movement in camera or emitter can cause large changes in depth comparisons, especially around shadow silhouettes. Inspired from noise-to-noise training \cite{NoiseToNoise}, we train our network to learn from pairs of noisy inputs, in addition to the traditional supervised learning pair. Our network intrinsically learns to damp sharp changes due to small perturbations with minimal impact on overall quality as shown in figure \ref{fig:effectOfPerturbationLoss}. At each training iteration, we perturb the camera and emitter position by a small value proportional to the distance from the scene and size of emitter. For each perturbation, we collect the input buffers for training. The target is evaluated for only one of the perturbations. We evaluate the network on each perturbations and minimize the differences in the perturbed outputs as an additional loss function. All network instances evaluating the perturbed inputs share the same weights while backpropagation is only enabled through one instance, as 
\begin{equation}
\mathcal{L} = L(\textcolor{red}{x_0}, \tilde{x}) + \textstyle\sum_{i=1}^p L(\textcolor{red}{x_0}, x_i), 
\label{eq:lossfunc}
\end{equation}
where $L(y, \tilde{y}) = \alpha \cdot |y - \tilde{y}| + (1 - \alpha) \cdot VGG19(y, \tilde{y})$, and $x_i$ and $\tilde{x}$ are the network outputs and target. Only one network output-$\textcolor{red}{x_0}$ has backpropagation enabled through it. We set $\alpha = 0.9$ and the number of perturbations $p = 3$.
\subsection{Temporal stability measurement}

 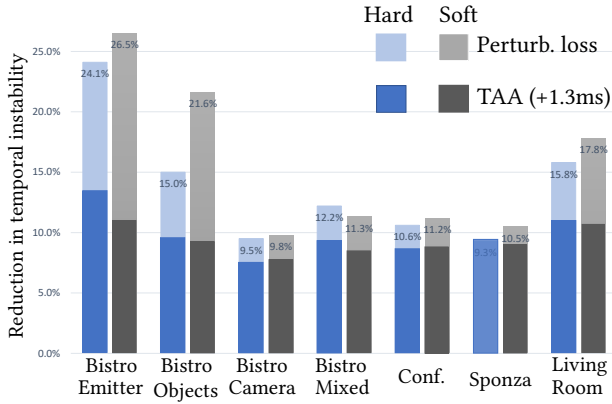
\begin{figure}[t!]
 \begin{tikzpicture}
    \node[anchor=south west,inner sep=0] at (0.0,0) {
    \includegraphics[width=12.0cm, trim={0cm 4.25cm 0cm 2.5cm},clip]{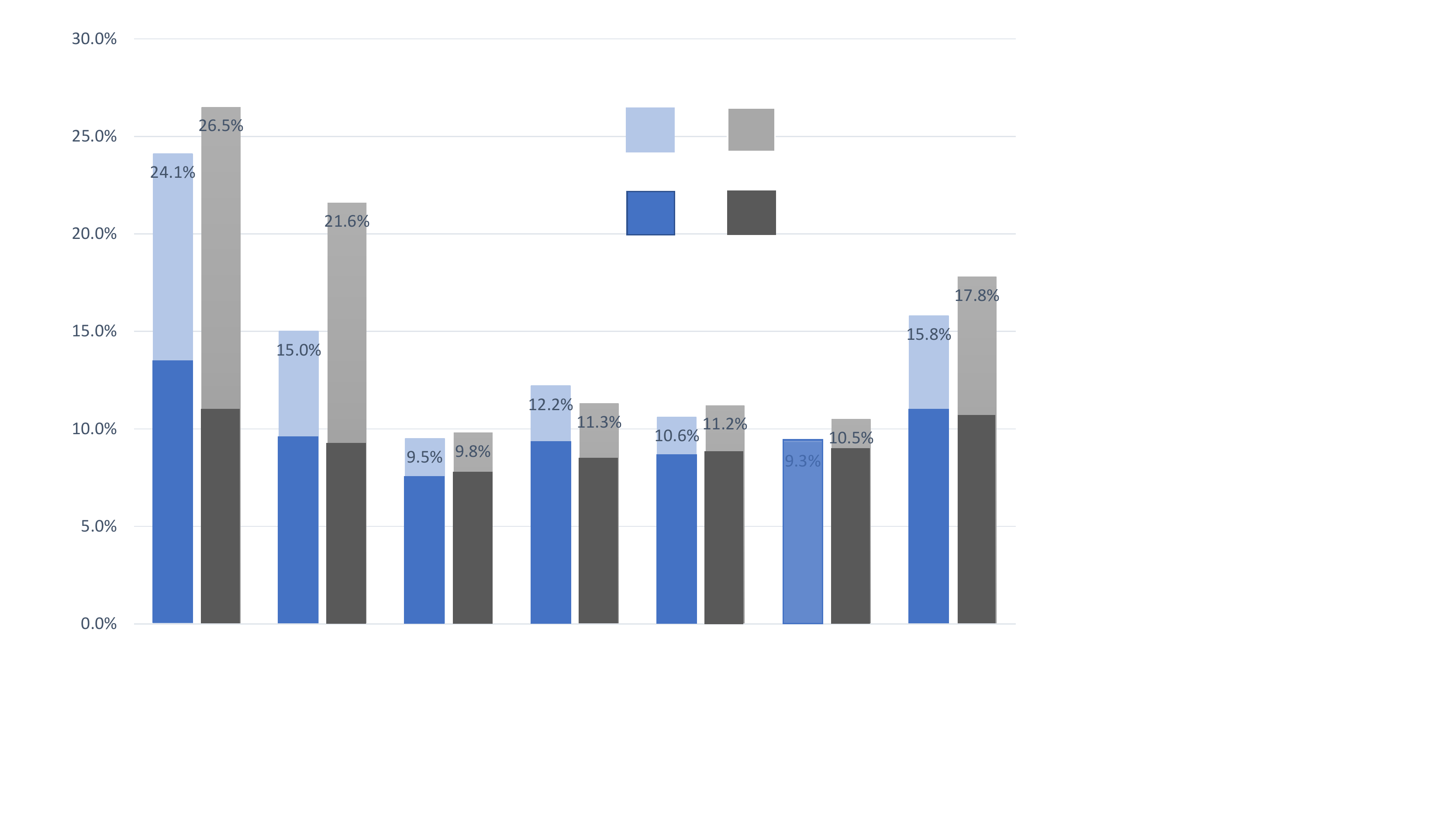}};
    \node[anchor=south west, rotate=90] at (0.65, 0.45- 0.25) { \textcolor{black}{\small Reduction in temporal instability} };
    \node[anchor=south west] at (4.9, 4.65- 0.25) { \textcolor{black}{Hard} };
    \node[anchor=south west] at (5.9, 4.65- 0.25) { \textcolor{black}{Soft} };
    \node[anchor=south west] at (6.4, 4.25- 0.25) { \textcolor{black}{Perturb. loss} };
    \node[anchor=south west] at (6.4, 3.5 - 0.25) { \textcolor{black}{TAA (+1.3ms)} };
    \node[anchor=south west] at (1.2, 0.0 - 0.25) { \textcolor{black}{\small Bistro} };
    \node[anchor=south west] at (1.1, -0.3 - 0.25) { \textcolor{black}{\small Emitter} };
    \node[anchor=south west] at (2.2, 0.0 - 0.25) { \textcolor{black}{\small Bistro} };
    \node[anchor=south west] at (2.1, -0.4 - 0.25) { \textcolor{black}{\small Objects} };
    \node[anchor=south west] at (3.2, 0.0 - 0.25) { \textcolor{black}{\small Bistro} };
    \node[anchor=south west] at (3.1, -0.3 - 0.25) { \textcolor{black}{\small Camera} };
    \node[anchor=south west] at (4.25, 0.0- 0.25) { \textcolor{black}{\small Bistro} };
    \node[anchor=south west] at (4.25, -0.3- 0.25) { \textcolor{black}{\small Mixed} };
    \node[anchor=south west] at (5.35, -0.15- 0.25) { \textcolor{black}{\small Conf.} };
    \node[anchor=south west] at (6.3, -0.25- 0.25) { \textcolor{black}{\small Sponza} };
    \node[anchor=south west] at (7.4, -0.05- 0.25) { \textcolor{black}{\small Living} };
    \node[anchor=south west] at (7.4, -0.3- 0.25) { \textcolor{black}{\small Room} };
    
\end{tikzpicture}
\caption{\label{fig:perturbLossCompare}Comparing reduction in temporal instability between \textit{Perturbation Loss} and TAA. Reference is trained without perturb. loss or TAA. TAA is implemented as an additional pass after network (trained without perturbation loss) evaluation requiring extra 1.3ms.}
\end{figure}

Several techniques exist for measuring perceptual similarity with respect to a single reference frame \cite{SSIM, FLIP} or reference video \cite{ForVideoVdp}. These techniques measure the spatio-temporal difference which indicates the overall reconstruction quality across screen-space and time. Since we sacrifice spatial quality for temporal stability, using these metric may not indicate a reduction in temporal instability due perturbation loss, even when there is a clear visual improvement in temporal stability. Thus we formulate our own metric to measure only the temporal changes without considering spatial similarity with reference. To measure flickering, we find the motion-vector adjusted per-pixel temporal difference~\cite{SurveyTAA}. Since flickering can be quantified as an abrupt change in the pixel intensities between frames, we penalize the large differences more by passing the temporal pixel difference through an exponential. We aggregate the result across all pixels and frames, with
\begin{equation}
E = \frac{1}{P}\sum_{p, t} \left\{\exp(\alpha D_t(p)) - 1\right\}~,
\end{equation}
where $D_t(p) = |I_t(p) - I_{t-1}(m(p))|$ is a per-pixel difference between two consecutive frames at time $t$ and $m(p)$ abstracts away the motion-vector adjusted lookup at pixel $p$ in the previous frame. We set $\alpha=3$, which controls the penalty for large changes in intensity through time, and the normalizing factor $P$ is the total number of pixels. We reject pixels that fail depth and normal comparison with its reprojection.

Figure \ref{fig:perturbLossCompare} contrasts the effect on temporal stability between our loss and TAA (1 last frame): the improvement in temporal stability with our perturbation loss is strongest in scenes with dynamic emitters and non-negligible, albeit smaller with dynamic view.

\begin{figure}[b!]
 \begin{tikzpicture}
    \node[anchor=south west,inner sep=0] at (0.0,0) {
    \includegraphics[width=18.5cm, trim={0cm 11.5cm 0cm 0.0cm},clip]{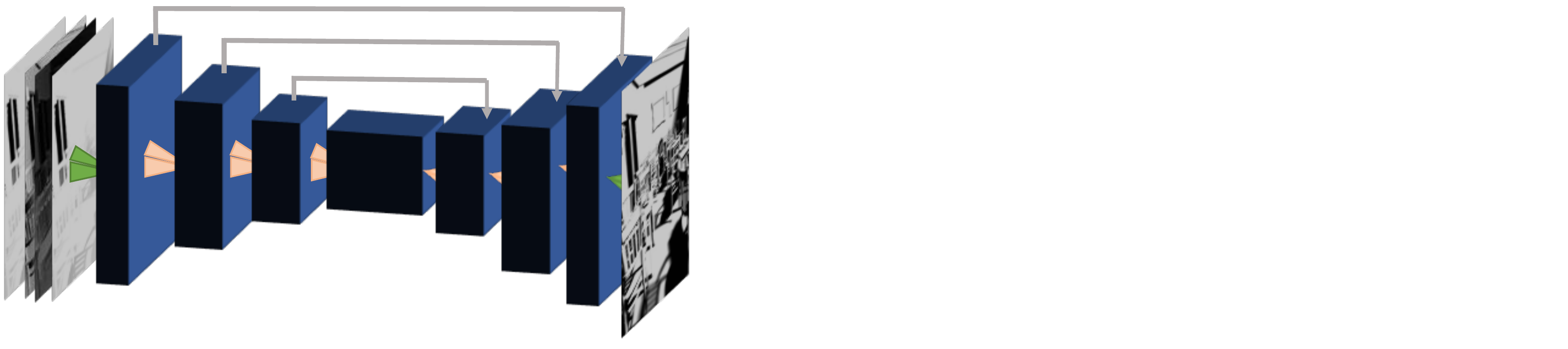}};
    \node[anchor=south west, rotate=90] at (0.4, 0.35-0.53) {
     \textcolor{black}{\small Time} };
     \node[anchor=south west, rotate=90] at (0.7, 0.45-0.53) { \textcolor{black}{\small (ms)} };
    \node[anchor=south west] at (0.6, 0.75-0.53) { 
     \textcolor{black}{\small 4.99} };
     \node[anchor=south west] at (0.6, 0.4-0.53) { \textcolor{OliveGreen}{\small 0.00 (Optimized)} };
      \node[anchor=south west] at (0.6, 0.4-0.85) { \textcolor{black}{\small Layer 0} };
    \node[anchor=south west] at (1.6, 1.2-0.53) { \textcolor{black}{\small 2.01} };
    \node[anchor=south west] at (1.6, 0.95-0.53) { \textcolor{OliveGreen}{\small 1.72} };
    \node[anchor=south west] at (2.55, 1.5-0.53) { \textcolor{black}{\small 0.91} };
    \node[anchor=south west] at (2.55, 1.25-0.53) { \textcolor{OliveGreen}{\small 0.93} };
    \node[anchor=south west] at (3.55, 1.75-0.53) { \textcolor{black}{\small 0.39} };
    \node[anchor=south west] at (3.55, 1.5-0.53) { \textcolor{OliveGreen}{\small 0.40} };
    \node[anchor=south west] at (4.5, 1.75-0.53) { \textcolor{black}{\small 0.57} };
    \node[anchor=south west] at (4.5, 1.5-0.53) { \textcolor{OliveGreen}{\small 0.59} };
     \node[anchor=south west] at (3.8, 1.5-0.95) { \textcolor{black}{\small Layer 4} };
    \node[anchor=south west] at (5.3, 1.5-0.53) { \textcolor{black}{\small 0.72} };
    \node[anchor=south west] at (5.3, 1.25-0.53) { \textcolor{OliveGreen}{\small 0.69} };
    \node[anchor=south west] at (6.0, 1.0-0.53) { \textcolor{black}{\small 1.67} };
    \node[anchor=south west] at (6.0, 0.75-0.53) { \textcolor{OliveGreen}{\small 1.32} };
     \node[anchor=south west] at (6.7, 0.6-0.53) { 
     \textcolor{black}{\small 5.98} };
     \node[anchor=south west] at (6.7, 0.3-0.53) { \textcolor{OliveGreen}{\small 0.24} };
      \node[anchor=south west] at (6.7, 0.4-0.95) { \textcolor{black}{\small Layer 0} };
     \node[anchor=south west] at (0, 3.8) { \textcolor{black}{\tiny $(1K,2K,4)$} };
     \node[anchor=south west, rotate=90] at (1.55, 1.4) { \textcolor{white}{\tiny $(512,1K,16)$} };
     \node[anchor=south west, rotate=90] at (2.55, 1.35) { \textcolor{white}{\tiny $(256,512,32)$} };
     \node[anchor=south west, rotate=90] at (3.5, 1.45) { \textcolor{white}{\tiny $(128,256,64)$} };
     \node[anchor=south west] at (3.8, 1.85) { \textcolor{white}{\tiny $(64,128,256)$} };
     \node[anchor=south west] at (7.1, 3.65) { \textcolor{black}{\tiny $(1K,2K,1)$} };
     
\end{tikzpicture}
\caption{\label{fig:netUnOpLayerTiming}Layer-wise performance optimization for a 1024$\times$2048 input.}
\end{figure}
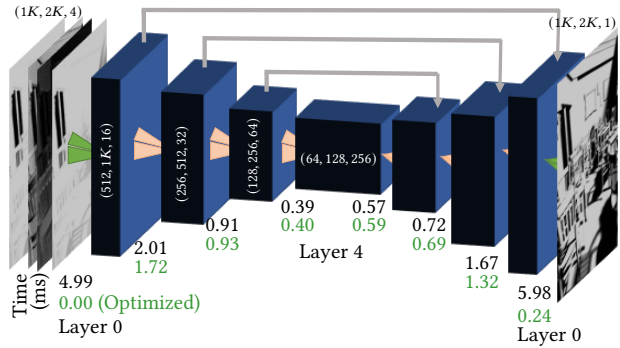
\subsection{Network architecture and optimizations \label{subsect:archOpti}}

The original UNet \cite{UNet15} architecture is too slow (>100ms) to fit into a real-time graphics pipeline. As such, we start with trimming the network down. Our generic network has 5 layers, however, each layer composed of one 3x3 convolution and one 1x1 convolution layers as opposed to the standard double 3x3 convolution. A major departure from the original UNet is using bi-linear interpolation instead of expensive transpose convolutions for upscaling. We also use algebraic sum instead of a concatenation layer for merging the skip connections on the decoder side. A positive side effect of using a sum layer is the reduction in the number of hidden units on the decoder side. With these modifications we reduce the network size from 25MB to just 2.5MB, while the runtime is minimized to 28ms. Quantizing the network to half-precision further reduces the size to 1.5MB and 17ms runtime.

Other modifications for improving temporal stability without affecting performance includes using Average-pool instead of Max-pool and removing the skip connection in the first layer. Replacing max-pool with average-pool reduces extremities during processing and smooths out the output. As raw shadow map depth values are prone to aliasing noise, removing the first skip connection ensures the noisy input does not affect the output directly.

\begin{figure}[b!]
\begin{tikzpicture}
    \node[anchor=south west,inner sep=0] at (0,4){
    \includegraphics[width=17.0cm, trim={0 11.25cm 0 0cm},clip]{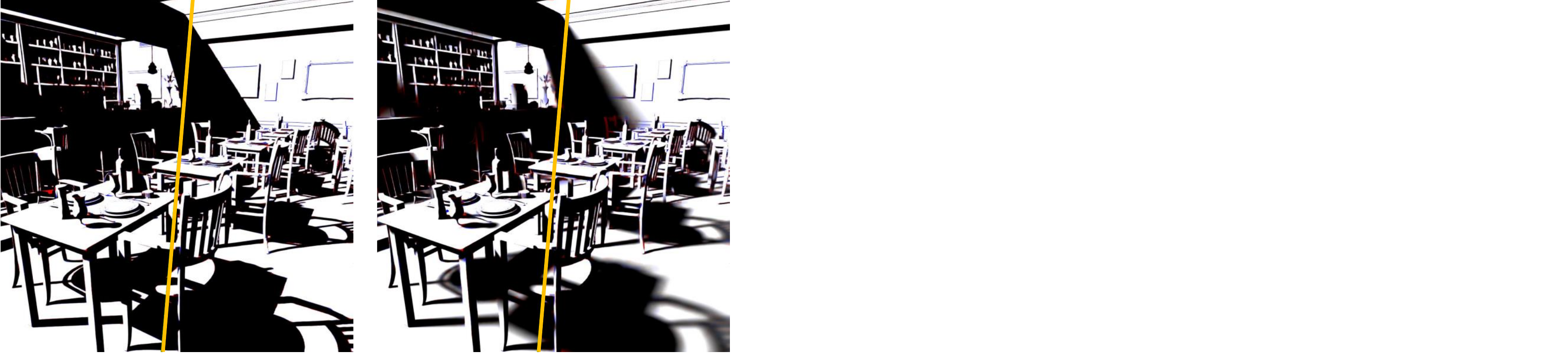}};
    \node[anchor=south west, rotate=90] at (0, 5) {\textcolor{black}{Hard-shadow} };
    \node[anchor=south west, rotate=90] at (4.15, 5) {\textcolor{black}{Soft-shadow} };
    \node[anchor=south west] at (0, 7.9) { \textcolor{black}{Before, 17.3ms} };
    \node[anchor=south west] at (2.1, 7.9) { \textcolor{black}{After, 5.84ms} };
    \node[anchor=south west] at (0.5, 7.5) { \textcolor{white}{\small MSE: 0.077 } };
    \node[anchor=south west] at (2.3, 7.55) { \textcolor{black}{\small MSE: 0.085} };
    \node[anchor=south west] at (4, 7.9) { \textcolor{black}{Before, 17.3ms} };
    \node[anchor=south west] at (6.1, 7.9) { \textcolor{black}{After, 5.84ms} };
    \node[anchor=south west] at (4.55, 7.5) { \textcolor{white}{\small MSE: 0.118 } };
    \node[anchor=south west] at (6.3, 7.55) { \textcolor{black}{\small MSE: 0.128} };
\end{tikzpicture}
\caption{Figure showing the effect of all optimizations in section ~\ref{subsect:archOpti}.}
\label{fig:beforeAfterOptimization}
\end{figure}
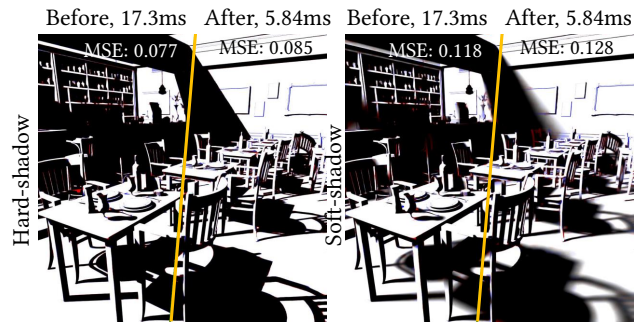

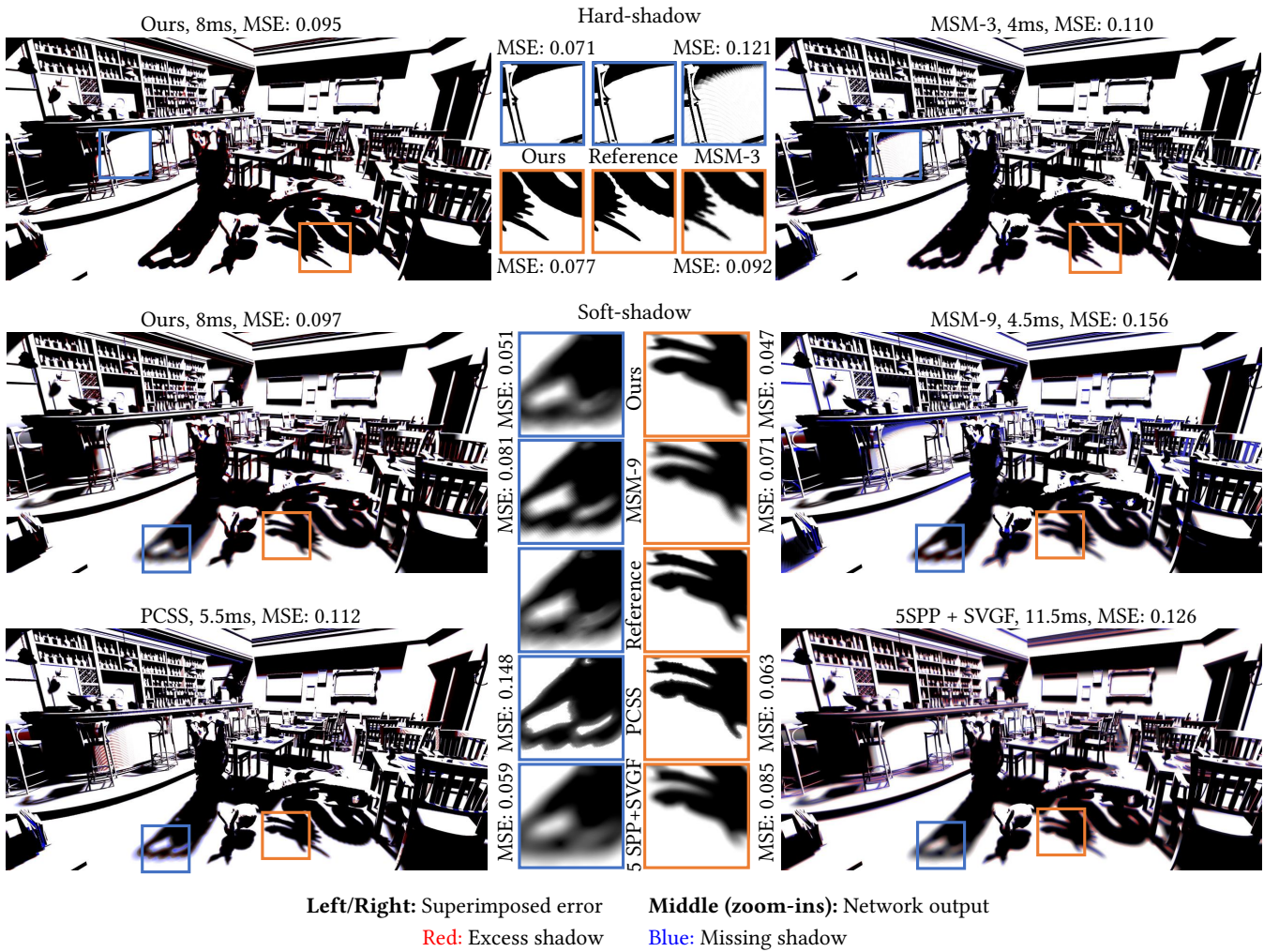
\begin{figure*}[t!]
\begin{tikzpicture}
    \node[anchor=south west,inner sep=0] at (0,4+4.3){
    \includegraphics[width=\textwidth,trim={0cm 10.9cm 0 0cm},clip]{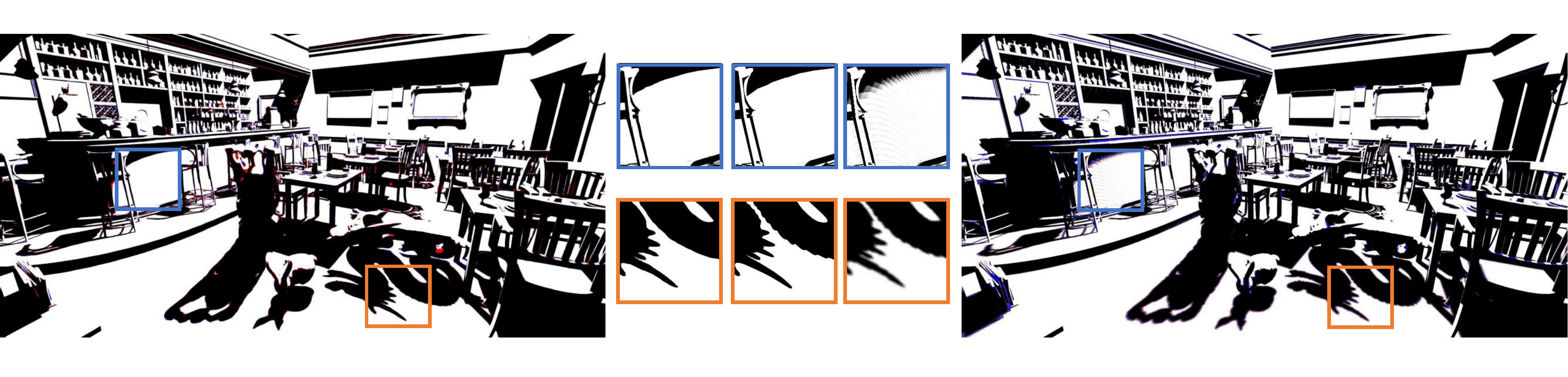}};
    
    \node[anchor=south west] at (1.8, 7.85 +4.3) {\textcolor{black}{Ours, 8ms, MSE: 0.095} };
    \node[anchor=south west] at (5.5+2.5, 8.0 + 4.3) {\textcolor{black}{Hard-shadow} };
    \node[anchor=south west] at (13, 7.85 + 4.3) {\textcolor{black}{MSM-3, 4ms, MSE: 0.110} };
    \node[anchor=south west] at (6.9-0.05, 7.55 + 4.3) {\textcolor{black}{MSE: 0.071} };
    \node[anchor=south west] at (9.4-0.05, 7.55 + 4.3) {\textcolor{black}{MSE: 0.121} };
    \node[anchor=south west] at (7.2, 6.0 + 4.3) {\textcolor{black}{Ours} };
    \node[anchor=south west] at (8.2-0.05, 6.0 + 4.3) {\textcolor{black}{Reference} };
    \node[anchor=south west] at (9.7-0.05, 6.0 + 4.3) {\textcolor{black}{MSM-3} };
    \node[anchor=south west] at (6.9-0.05, 4.45 + 4.3) {\textcolor{black}{MSE: 0.077} };
    \node[anchor=south west] at (9.4-0.05, 4.45 + 4.3) {\textcolor{black}{MSE: 0.092} };
    
     \node[anchor=south west,inner sep=0] at (0,8-8){
    \includegraphics[width=\textwidth,trim={0cm 3.0cm 0 0cm},clip]{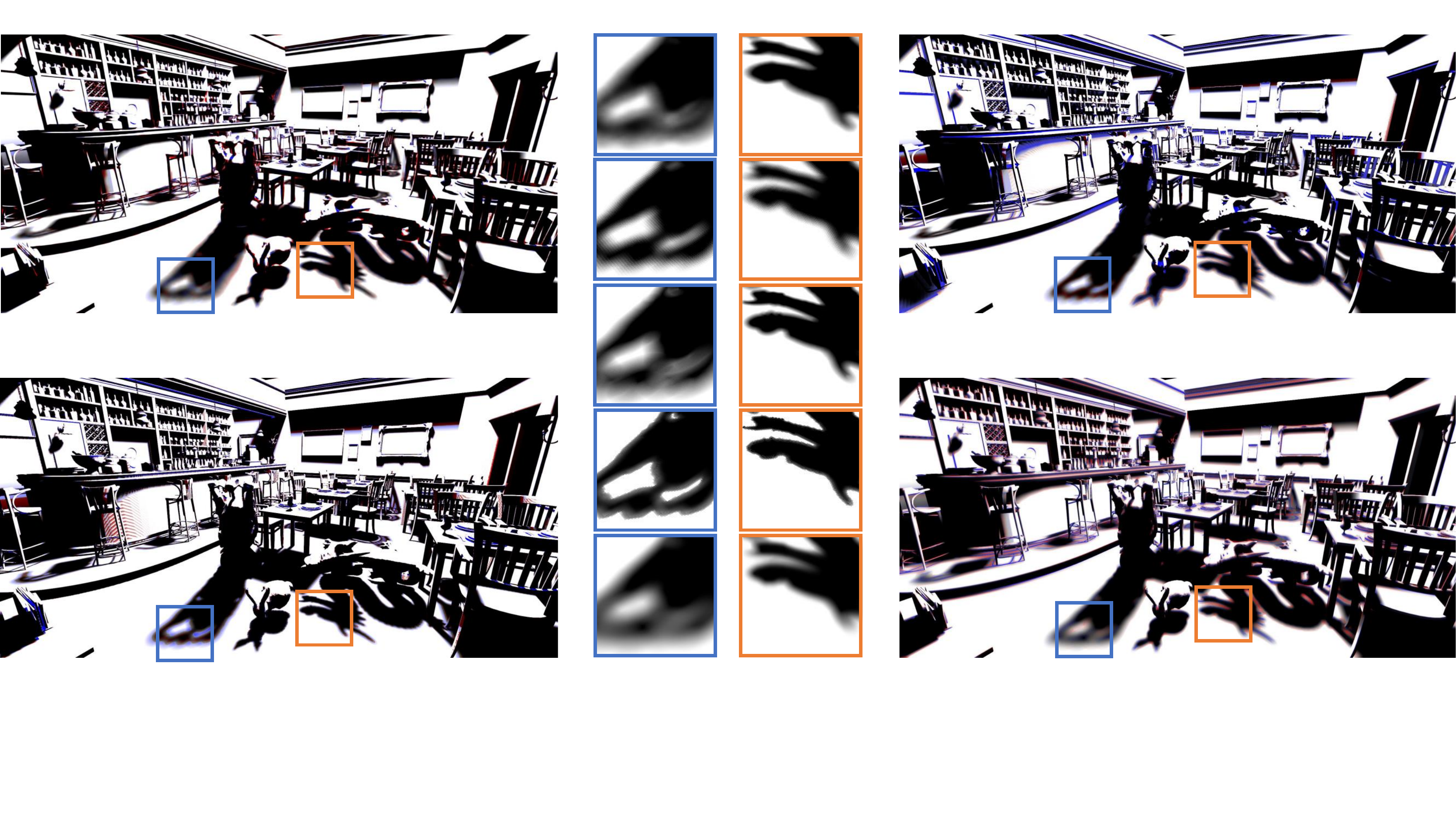}};
    \node[anchor=south west] at (1.8, 15.95 - 8) {\textcolor{black}{Ours, 8ms, MSE: 0.097} };
    \node[anchor=south west] at (5.5 +2.5, 16.1 - 8) {\textcolor{black}{Soft-shadow } };
    \node[anchor=south west] at (13, 15.95 - 8) {\textcolor{black}{MSM-9, 4.5ms, MSE: 0.156} };
    \node[anchor=south west] at (1.8, 11.75 - 8) {\textcolor{black}{PCSS, 5.5ms, MSE: 0.112} };
    \node[anchor=south west] at (12.5, 11.75 - 8) {\textcolor{black}{5SPP + SVGF, 11.5ms, MSE: 0.126} };
    \node[anchor=south west, rotate=90] at (9.17-0.05, 14.8 - 8) {\textcolor{black}{Ours} };
    \node[anchor=south west, rotate=90] at (9.17-0.05, 13.2 - 8) {\textcolor{black}{MSM-9} };
    \node[anchor=south west, rotate=90] at (9.17-0.05, 11.4 - 8) {\textcolor{black}{Reference} };
     \node[anchor=south west, rotate=90] at (9.17-0.05, 10.2 - 8) {\textcolor{black}{PCSS} };
     \node[anchor=south west, rotate=90] at (9.17-0.05, 8.25 - 8) {\textcolor{black}{5 SPP+SVGF} };
      \node[anchor=south west, rotate=90] at (7.35-0.05, 14.55 - 8) {\textcolor{black}{MSE: 0.051} };
    \node[anchor=south west, rotate=90] at (7.35-0.05, 13.05 - 8) {\textcolor{black}{MSE: 0.081} };
    \node[anchor=south west, rotate=90] at (7.35-0.05, 9.95 - 8) {\textcolor{black}{MSE: 0.148} };
     \node[anchor=south west, rotate=90] at (7.35-0.05, 8.4 - 8) {\textcolor{black}{MSE: 0.059} };
     \node[anchor=south west, rotate=90] at (11.0, 14.55 - 8) {\textcolor{black}{MSE: 0.047} };
    \node[anchor=south west, rotate=90] at (11.0, 13.05 - 8) {\textcolor{black}{MSE: 0.071} };
    \node[anchor=south west, rotate=90] at (11.0, 9.95 - 8) {\textcolor{black}{MSE: 0.063} };
     \node[anchor=south west, rotate=90] at (11.0, 8.4 - 8) {\textcolor{black}{MSE: 0.085} };
    \node[anchor=south west] at (4.15, -0.40) { \textcolor{black}{\textbf{Left/Right: }Superimposed error} };
    \node[anchor=south west] at (9.0, -0.40) { \textcolor{black}{\textbf{Middle (zoom-ins): }Network output} };
    \node[anchor=south west] at (5.8, -0.78) { \textcolor{red}{Red:} \textcolor{black}{Excess shadow} };
    \node[anchor=south west] at (9.0, -0.85) { \textcolor{blue}{Blue:} \textcolor{black}{Missing shadow} };
\end{tikzpicture}
\caption{\label{fig:comparisonUnseenObjects} Comparing our network's ability to generalize to unseen objects (\textsc{buddha}, \textsc{bunny}, \textsc{dragon}) with other competing techniques (MSM, PCSS, Raytracing \& Denoising) for hard and soft shadows. MSM-3 and MSM-9 are Moment Shadow Map variants using 3$\times$3 and 9$\times$9 prefiltering kernels.}
\end{figure*}

At this stage, we analyze the performance and error of our network before optimizing it further. Our validation error is $6.67 \times 10^{-3}$ over an ensemble test scenes. From figure \ref{fig:netUnOpLayerTiming}, we see that the first layer (combined encoder-decoder) requires more time compared to the rest of the layers combined. Moving from inner (\#4) to outer layers (\#0), the resolution is quadrupled while the number of channels is halved. Consequently, the effective cached memory bandwidth doubles as we move from inner to outer layers; however, with increasing resolution, memory operations are also more prone to cache misses. In practice, we see more than doubling of runtime as we move towards the outer layers. Refer supplemental section 3.

We further optimize by changing the first layer which consumes disproportionately more time. A naive approach is to replace the first layer with a downsampler on the encoder side and upsampler on the decoder side of UNet . However, simple downsampling and upsampling loses information contained in the input and also produces less sharp output. Instead, we flatten a 2x2 square of pixels into 4 separate channels and use the restructured buffer as the input for second layer. Thus we rearrange the input and change the buffer dimensions from  ($h\times w\times ch$) to ($h/2\times w/2\times 4ch$). More concretely, instead of feeding the first layer with full resolution ($1024\times2048$) input with 4 channels, we feed the second layer directly with quarter resolution ($512\times1024$) input with 16 channels. We do the inverse on the decoder side; rearrange 4 output channels into a 2x2 pixel square. Note that the rearrangement of buffers does not add any extra temporary storage for the second layer while removing the first layer (Conv2D operations) completely. On the decoder side, to improve training convergence, we upscale the first output channel to full resolution using a bi-linear interpolation. We then add rest of the three channels to the interpolated output, filling in rest of the details. The performance of our optimized network is 5.8ms.

\subsection{Network depth optimizations}

\begin{figure}[t!]
\begin{tikzpicture}
    \node[anchor=south west,inner sep=0] at (0,0){
    \includegraphics[width=11.5cm, trim={0cm 0.5cm 0 0cm},clip]{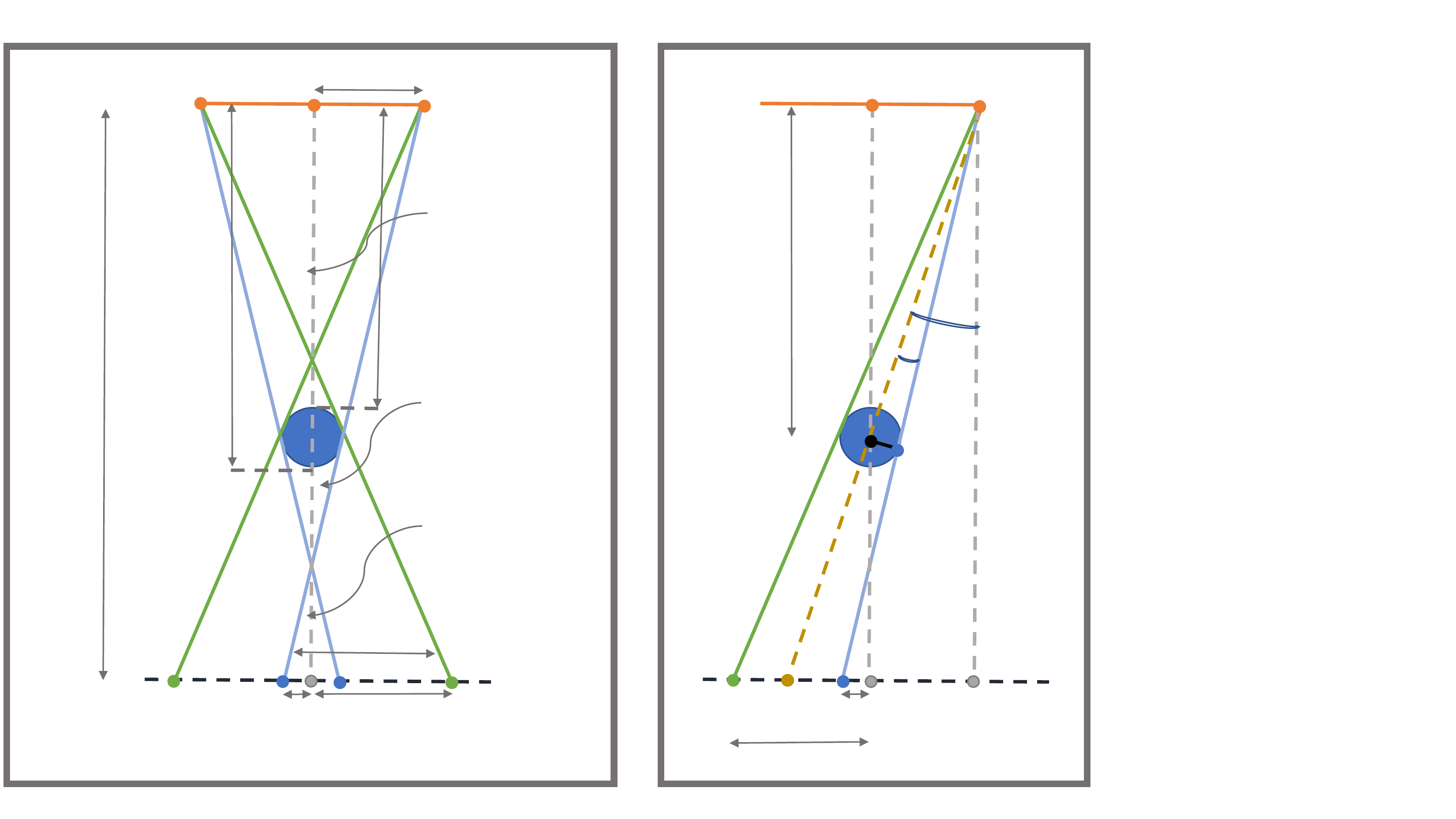}};
    \node[anchor=south west] at (0.75, 5.95) { \textcolor{black}{\small a. Model estimating penumbra}};
    \node[anchor=south west] at (2.75, 5.525) { \textcolor{black}{$r_e$}};
    \node[anchor=south west] at (1.0, 5.45) { \textcolor{black}{Emitter}};
    \node[anchor=south west] at (3.25, 4.45) { \textcolor{black}{Center line}};
    \node[anchor=south west] at (2.95, 3.525) { \textcolor{black}{$z_{min}$}};
    \node[anchor=south west] at (1.05, 3.75) { \textcolor{black}{$z_{max}$}};
    \node[anchor=south west] at (0.4, 2.8) { \textcolor{black}{$z_f$}};
    \node[anchor=south west] at (3.25, 2.9) { \textcolor{black}{$x_a < 0$}};
    \node[anchor=south west] at (3.25, 1.9) { \textcolor{black}{$x_a > 0$}};
    \node[anchor=south west] at (3.4, 1.3) { \textcolor{black}{Penumbra}};
    \node[anchor=south west] at (3.4, 0.95) { \textcolor{black}{$\approx x_a + x_b$}};
    \node[anchor=south west] at (2.1, 0.45) { \textcolor{black}{$x_a$}};
     \node[anchor=south west] at (3.0, 0.40) { \textcolor{black}{$x_b$}};
     
     \node[anchor=south west] at (5.35, 5.95) { \textcolor{black}{\small b. Estimating parameters}};
     \node[anchor=south west] at (6.7, 5.45) { \textcolor{black}{\small P}};
     \node[anchor=south west] at (7.5, 5.45) { \textcolor{black}{\small A}};
     \node[anchor=south west] at (5.7, 4.0) { \textcolor{black}{$z_m$}};
     \node[anchor=south west] at (7.3, 3.35) { \textcolor{black}{$\theta$}};
     \node[anchor=south west] at (6.9, 2.95) { \textcolor{black}{$\theta_{\delta}$}};
     \node[anchor=south west] at (6.5, 2.65) { \textcolor{black}{\small M}};
     \node[anchor=south west] at (7.05, 2.55) { \textcolor{black}{\small D}};
      \node[anchor=south west] at (6.05, 0.5) { \textcolor{black}{\small B}};
     \node[anchor=south west] at (6.9, 0.5) { \textcolor{black}{\small Q}};
     \node[anchor=south west] at (6.55, 0.45) { \textcolor{black}{$x_a$}};
      \node[anchor=south west] at (7.5, 0.5) { \textcolor{black}{\small C}};
      \node[anchor=south west] at (6.05, 0.05) { \textcolor{black}{$x_b$}};
\end{tikzpicture}
\caption{\label{fig:estimatePenubraWidth} A simplified model to estimate the penumbra size at a given pixel. We assume our occluder is spherical in shape, forming a convex bounding-sphere around the occlusion geometry as shown in figure (a) on the left. Figure (b) shows a simplified diagram to estimate the parameters $\theta, \theta_\delta$.}
\end{figure}
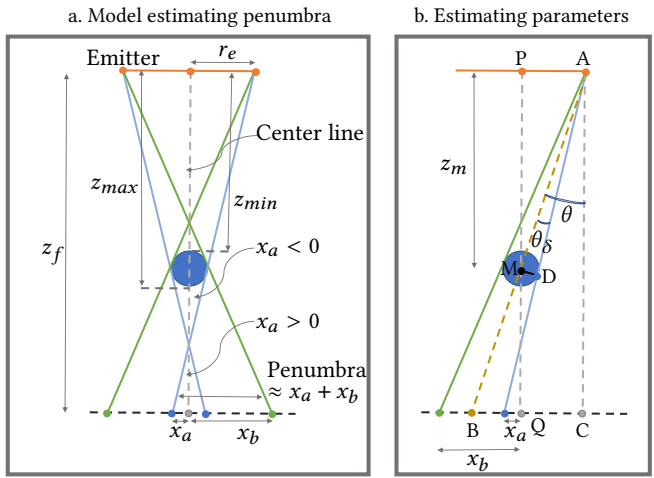

Our optimizations so far are generic and apply across scene and emitter configurations. Below, we explore \textit{scene specific} optimizations and tune our network architecture for compactness. Shallower networks have many pragmatic benefits: it has exponentially (power of 2) fewer parameters, is faster to train, and admits faster runtime inference. Instead of relying on adhoc architecture tuning, we will choose architectures based on their ability to capture the shadowing effects we target. Specifically, we will build a simple model to estimate the maximum penumbra size for a scene configuration, and then relate this size exactly to the depth of the network suited to reproducing them. We empirically validate our model.

To compute the world space penumbra size, our simplified model assumes a spherical occluder (or, conservatively, a bounding sphere around occluding geometry). When generating training samples, we additionally measure the minimum and maximum occluder distances $z_{max}$, $z_{min}$ (figure \ref{fig:estimatePenubraWidth}, a). We then estimate the penumbra width at a pixel as the sum of inner ($x_a$) and outer ($x_b$) penumbra (figure \ref{fig:estimatePenubraWidth}, b) as $\{x_a,x_b\} = z_f\,\tan(\theta \pm \theta_{\delta}) - r_e$. We derive the parameters $\theta, \theta_\delta$ in the supplemental, section 4.

\begin{table}[b!]
\caption{\label{tab:empericalDepthVerification} MSE for variable penumbra sizes and with 3/5/7-layer nets.}
\begin{tabular}{|c|ccc|}
\hline
\multirow{2}{*}{\begin{tabular}[c]{@{}c@{}}Predicted Penumbra\\ Width (in pixels)\end{tabular}} & \multicolumn{3}{c|}{\begin{tabular}[c]{@{}c@{}}Network layers $\rightarrow$\\ Receptive field size (in pixels)\end{tabular}} \\ \cline{2-4} 
 & \multicolumn{1}{c|}{\begin{tabular}[c]{@{}c@{}}3 $\rightarrow$ 24\end{tabular}} & \multicolumn{1}{c|}{\begin{tabular}[c]{@{}c@{}}5 $\rightarrow$ 96\end{tabular}} & \begin{tabular}[c]{@{}c@{}}7 $\rightarrow$ 384\end{tabular} \\ \hline
21 & \multicolumn{1}{c|}{0.005} & \multicolumn{1}{c|}{0.005} & 0.012 \\ \hline
90 & \multicolumn{1}{c|}{\textcolor{red}{0.083}} & \multicolumn{1}{c|}{0.009} & 0.018 \\ \hline
180 & \multicolumn{1}{c|}{\textcolor{red}{0.161}} & \multicolumn{1}{c|}{\textcolor{red}{0.042}} & 0.019 \\ \hline
\end{tabular}
\end{table}

After computing a histogram of penumbra sizes in screen-space for each pixel across all the training frames, we select the highest 95th percentile penumbra size as a conservative bound on the \textit{receptive field} size requirements for our neural architecture.

We can modulate the per-layer convolutional layer parameters (kernel size, stride) and pooling operation parameters in order to meet the target receptive field requirements. If we set each convolutional layers to halve the spatial resolution, the \textit{effective receptive field} of the network grows with $\times 2^{l}$ for an $l$-layer network. Exclusively using $3\times 3$ kernels, we can solve for $l = \log_2(p_w / 3)$, where $p_w$ is the conservative screen-space penumbra \textit{width}.

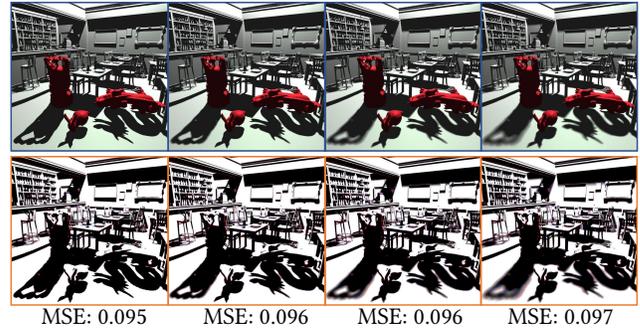
\begin{figure}[t!]
\begin{tikzpicture}
    \node[anchor=south west,inner sep=0] at (0,0){
    \includegraphics[width=8.5cm, trim={0cm 2.85cm 0 0cm},clip]{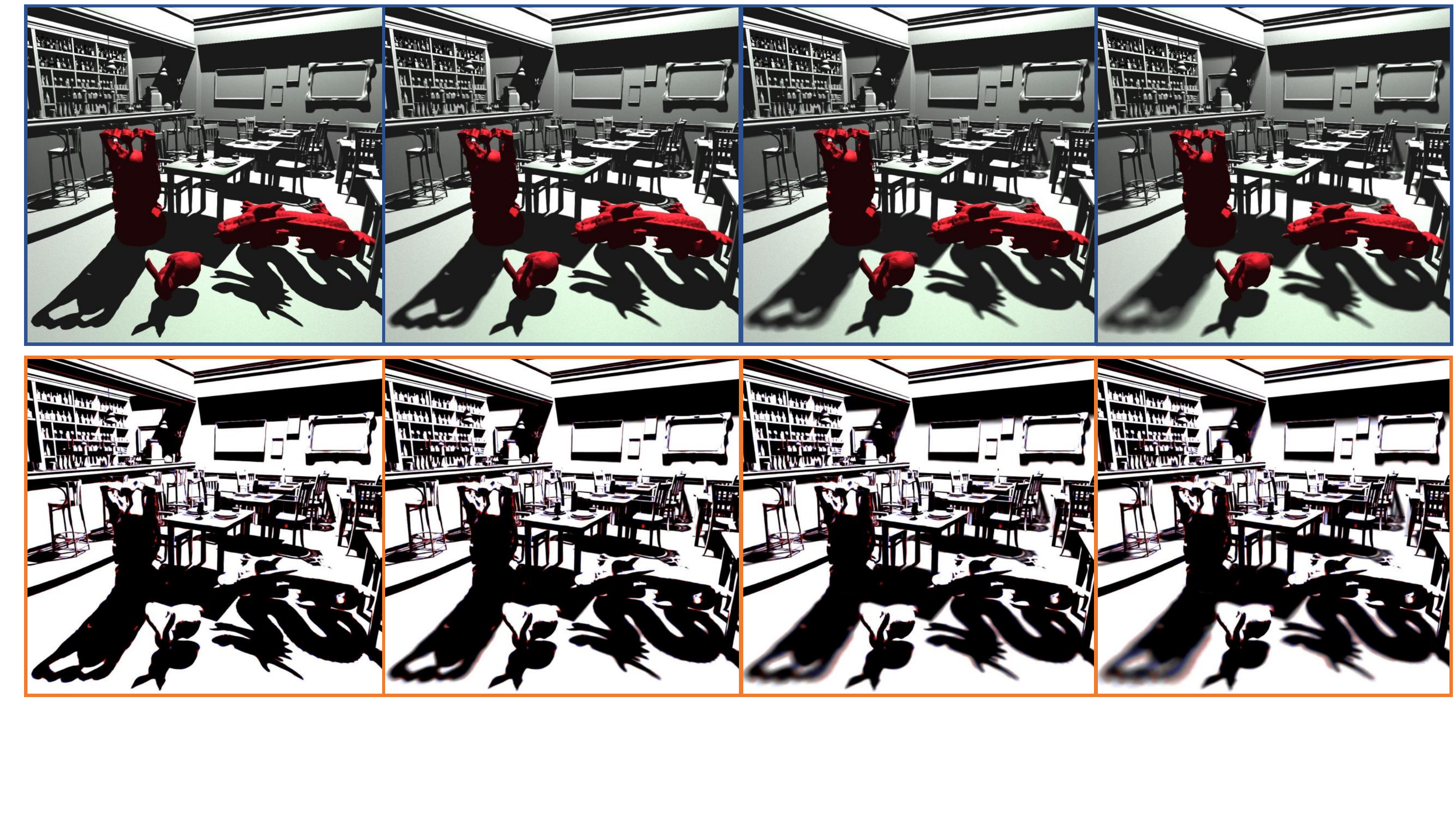}};
    \node[anchor=south west] at (0.45, 0.15-0.5) { \textcolor{black}{MSE: 0.095} };
    \node[anchor=south west] at (2.6, 0.15-0.5) { \textcolor{black}{MSE: 0.096} };
    \node[anchor=south west] at (4.65, 0.15-0.5) { \textcolor{black}{MSE: 0.096} };
    \node[anchor=south west] at (6.65, 0.15-0.5) {\textcolor{black}{MSE: 0.097} };
\end{tikzpicture}
\caption{\label{fig:resultsUntrainedObjects} Generalization to untrained objects (red, first row) in a trained scene. Second row visualizes false color error w.r.t.\ reference. From left to right, emitter size increases linearly from 0 (hard shadow) to 50 cm diameter.}
\end{figure}

Table \ref{tab:empericalDepthVerification} provides an empirical validation of our technique. We train three networks with 3,5, and 7 layers using the same dataset. The dataset consist of scenes with mixture of penumbra sizes. The penumbra size estimates are computed using our model. During inference, networks with \textit{receptive field} lower than the predicted penumbra size perform poorly as marked in \textcolor{red}{red} color. A more detailed analysis of the empirical verification is provided in the supplemental section 4.1.

\section{Results and comparisons}

We demonstrate our method on a diversity of scenarios. We augment static environments (e.g., rooms) included in our training set to include untrained objects at runtime, illustrating an important use case for interactive settings like games (figure \ref{fig:resultsUntrainedObjects}). We train a single network on the \textsc{Bistro interior} scene with varying emitter sizes, emitter positions and camera trajectories and introduce the (untrained objects) \textsc{Buddha}, \textsc{bunny}, and \textsc{dragon} for validation.

\paragraph*{Comparison.}
In the soft-shadowing regime, our comparisons focus on two classes of baseline methods: first, high-performance rasterization-based approximations such as Moment Shadow Maps (MSM) \cite{MSM15} and Percentage Closer Soft Shadows (PCSS) \cite{PCSS05} that align with our engineering (i.e., fully rasterization-based; no ray-tracing) and performance targets as our primary baseline; second, we use interactive GPU ray-tracing with post-process denoising as a more accurate ``interactive'' baseline, i.e., 5-SPP raytracing with SVGF~\cite{SVGF17}. Note that, unlike our method, neither MSM nor PCSS allow explicit control of penumbra style using emitter size; as such, we adjust the pre-filtering kernel size for MSM and PCSS to achieve a penumbra size that most closely matches reference renderings. We use kernel sizes of 3$\times$3 for MSM and 9$\times$9 for PCSS. Our PCSS baseline also includes a depth-aware post-filtering.

Our method consistently improves shadow quality at competitive performances (figure \ref{fig:comparisonUnseenObjects}). Refer to our video to observe the temporal stability of our results. For hard shadows, we compare to $3\times 3$ MSM, obtaining alias-free shadows without any light leaking. Please refer to the supplemental section 7 for more results, comparisons.

Runtime comparisons are measured at a resolution of 2k$\times$1k on an \textsc{AMD 5600X CPU} and \textsc{Nvidia 2080Ti} GPU. The timings in all figures, both main paper and supplemental, exclude G-Buffer generation which consistently requires an additional 2-3 ms (depending on the scene) across all techniques. Each scene is trained on $\leq 400$ images of resolution 2k$\times$1k on a cluster for roughly 16 hours (75 training epochs).

\section{Limitations}
\begin{figure}[b!]
\begin{tikzpicture}
    \node[anchor=south west,inner sep=0] at (0,0){
    \includegraphics[width=9.1cm, trim={0cm 11.25cm 0 0cm},clip]{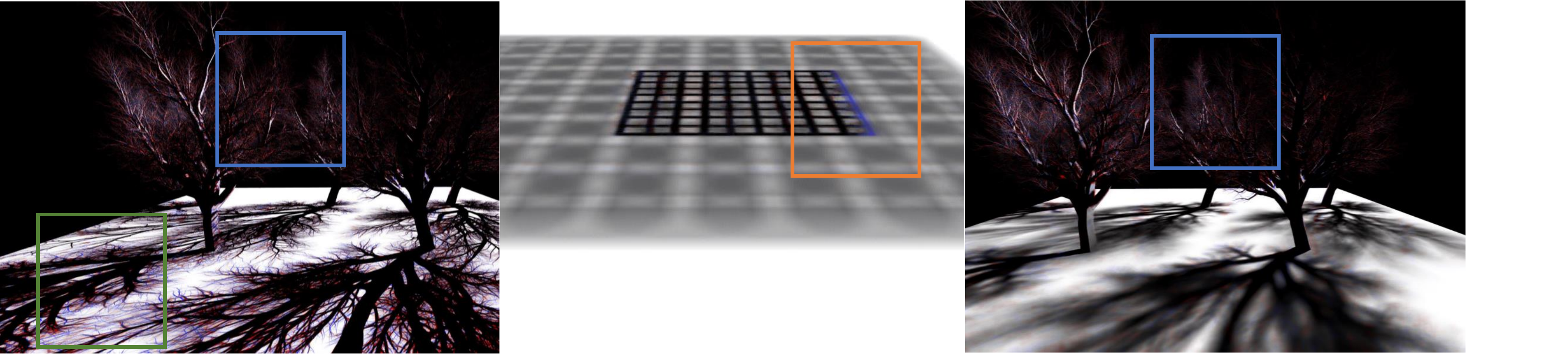}};
    \node[anchor=south west] at (0.5, 4.95 - 3) { \textcolor{black}{Foliage-hard} };
    \node[anchor=south west] at (2.9, 4.95 - 3) { \textcolor{black}{Overlapping Mesh} };
    \node[anchor=south west] at (6.2, 4.95 - 3) { \textcolor{black}{Foliage-soft} };
    \node[anchor=south west] at (0.5, 2.55 - 2.9) { \textcolor{black}{MSE :0.047} };
    \node[anchor=south west] at (3.4, 2.55 - 2.9) { \textcolor{black}{MSE: 0.003} };
    \node[anchor=south west] at (6.2, 2.55 - 2.9) { \textcolor{black}{MSE: 0.022} };
\end{tikzpicture}
\caption{\label{fig:limitations} Limitations of our method with high-frequency/high-depth complexity. Orange, blue, and green boxes highlight visual artifacts due to limited emitter and camera depth information, and due to undersampled training scenarios (i.e., loss of fine details).}
\end{figure}
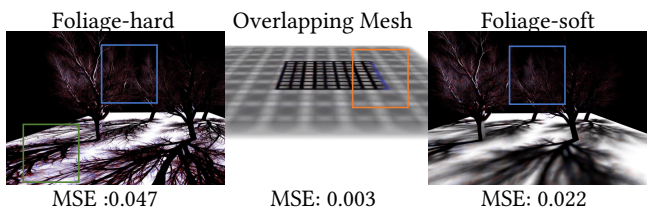

Our technique shares similar limitations to other screen-space methods. The unavailability of layered depth information, both in camera-~\cite{SSAO09} and emitter-space~\cite{SSSSGpuPro360} leads to an ill-posedness of the problem that results in approximation error. In camera space, the lack of peeled-depth data complicates the determination of mutual visibility between pixels. Similarly, computing the blur kernel size to soften shadow silhouettes relies on the distance between shading points and occluders, which is also unavailable in our setting. Our network and training methodology are effectively designed to compensate for this ill-posedness, bridging the visual gap in a diversity of object/scene arrangements by leveraging complex patterns inherent in the data. Figure ~\ref{fig:limitations} highlights a standard failure case and our supplemental includes additional discussion (section 8).

\section{Conclusions}

We presented a compact, fast neural method and training loss suited to temporally-coherent hard and soft shadows synthesis using only basic shadow map rasterized inputs. We showed that -- with a careful, problem-specific architecture design and a new, simple temporal loss -- a single small network can learn to hallucinate hard and soft shadows from varying emitter sizes and for a diversity of scenes. It is robust to the insertion of unseen objects, requires only a modest training budget, and precludes the need for any post-process denoising and/or TAA. Our approach yields stable hard and soft shadows with performance similar to workhorse interactive approximations and higher quality than (more expensive) GPU-raytracing and denoising (and TAA) alternatives.

Rasterization-based approaches for soft and hard shadows rely on heuristics and brittle manual tuning to achieve consistent, visually-desirable results. Our data-driven approach precludes such tuning, improving shadow quality at a modest cost, producing plausible and temporally-coherent soft shadows without any ray-tracing. Ours is a compact neural shading-based framework ~\cite{DeepShading16} suitable for low-power tiled-rendering systems, striking an interesting trade-off in a complex design space. We demonstrate benefits that largely offset the added training and integration complexity.

In the future, pursuing more aggressive neural architecture optimizations, including quantization and procedural architecture search, could likely further improve inference performance. When coupled with sparsification using, e.g., \textit{lottery ticket-based} method~\cite{lotteryTicketML}, we suspect that significant additional performance gains are possible, all without sacrificing quality.

\begin{acks}
We thank the reviewers for their constructive feedback, the ORCA for the \textsc{Amazon Lumberyard Bistro} model~\cite{ORCAAmazonBistro}, the Stanford CG Lab for the \textsc{Bunny}, \textsc{Buddha}, and \textsc{Dragon} models, Marko Dabrovic for the \textsc{Sponza} model and Morgan McGuire for the \textsc{Bistro}, \textsc{Conference} and \textsc{Living Room} models~\cite{MorganMcGuireArchive}. This work was done when Sayantan was an intern at Meta Reality Labs Research. While at McGill University, he was also supported by a Ph.D.\ scholarship from the \textit{Fonds de recherche du Québec -- nature et technologies}.
\end{acks}

\bibliographystyle{ACM-Reference-Format}
\bibliography{main}


\setcounter{section}{0}
\setcounter{equation}{0}
\setcounter{figure}{0}
\setcounter{table}{0}
\clearpage
\textbf{\huge Neural Shadow Mapping - Supplemental}
\vspace{100pt}
\section{Feature engineering}
\subsubsection{Overfitting and underfitting} Neural networks are known for their ability to fit a dataset, even when there is no underlying relationship between the features and the target - a phenomenon known as overfitting. An extreme example is mapping our ray-traced targets to white noise textures as features. In this case, the neural network acts as a hash-map, mapping input to the output. This might work for a very small dataset but will obviously fail for any larger dataset. In practice, some features provide crucial information and some do not. Thus, to isolate noise from information, we need a diverse dataset. In our case, we can create diversity by collecting data across various emitter and camera positions, and scenes. In fact, as we generate more training data, we run into the regime of underfitting, where the network lacks enough capacity to learn the details in the dataset.

\begin{table}[b!]
\caption{\label{tab:featuresearch} Validation error for various combination of buffers. At each row, we add 2 new features and remove features with sensitivity below 1.5\%. $+/-$ indicates set addition and subtraction.}
\centering
\begin{tabular}{|c|c|c|c|c|c|}
\hline
\multicolumn{3}{|c|}{Before pruning} & \multicolumn{3}{c|}{After pruning} \\ \hline
\begin{tabular}[c]{@{}c@{}}Fea.\\ Count\end{tabular} & \begin{tabular}[c]{@{}c@{}}Feature\\ Description\end{tabular} & \begin{tabular}[c]{@{}c@{}}Err\\ $10^{-3}$\end{tabular} & \begin{tabular}[c]{@{}c@{}}Fea.\\ Count\end{tabular} & \begin{tabular}[c]{@{}c@{}}Feature\\ Description\end{tabular} & \begin{tabular}[c]{@{}c@{}}Err\\ $10^{-3}$\end{tabular} \\ \hline
2 & $\left\{z/z_f, c_e\right\}$ & $6.86$ & \multicolumn{3}{c|}{No change} \\ \hline
4 & $+ \left\{z - z_f, c_c/d\right\}$ & $6.65$ & \multicolumn{3}{c|}{No change} \\ \hline
6 & $+ \left\{ \mathbf{n} \cdot \mathbf{n_e}, d\right\}$ & $6.61$ & 4 & $- \left\{\mathbf{n} \cdot \mathbf{n_e}, c_c/d\right\}$ & $6.69$ \\ \hline
\end{tabular}
\vspace{50pt}
\end{table}

\subsubsection{Feature selection}

As described in the main paper, we use \textit{sensitivity} as our metric to select and prune our features from the set $U = \left\{d, \mathbf{n}, z, \mathbf{n_e}, z_f, c_e, c_c \right\} + \left\{z - z_f, z / z_f, c_c / d, \mathbf{n} \cdot \mathbf{n_e}\right\}$. However, pruning based on sensitivity alone does not lead to a unique combination of buffers. There are sever combinations with similar sensitivity. We use validation error as tie-breaking rule in such scenario. For example as shown in table \ref{tab:featuresearch}, buffer combinations $\{z/z_f, z-z_f, c_e, c_c/d\}$ and  $\{z/z_f, z-z_f, \mathbf{n} \cdot \mathbf{n}, d\}$ have similar sensitivity but one has lower validation error.

\section{Temporal loss}

\begin{figure}[h]
\begin{tikzpicture}
    \node[anchor=south west,inner sep=0] at (0,0){
    \includegraphics[width=10.5cm, trim={0cm 0cm 1 0cm},clip]{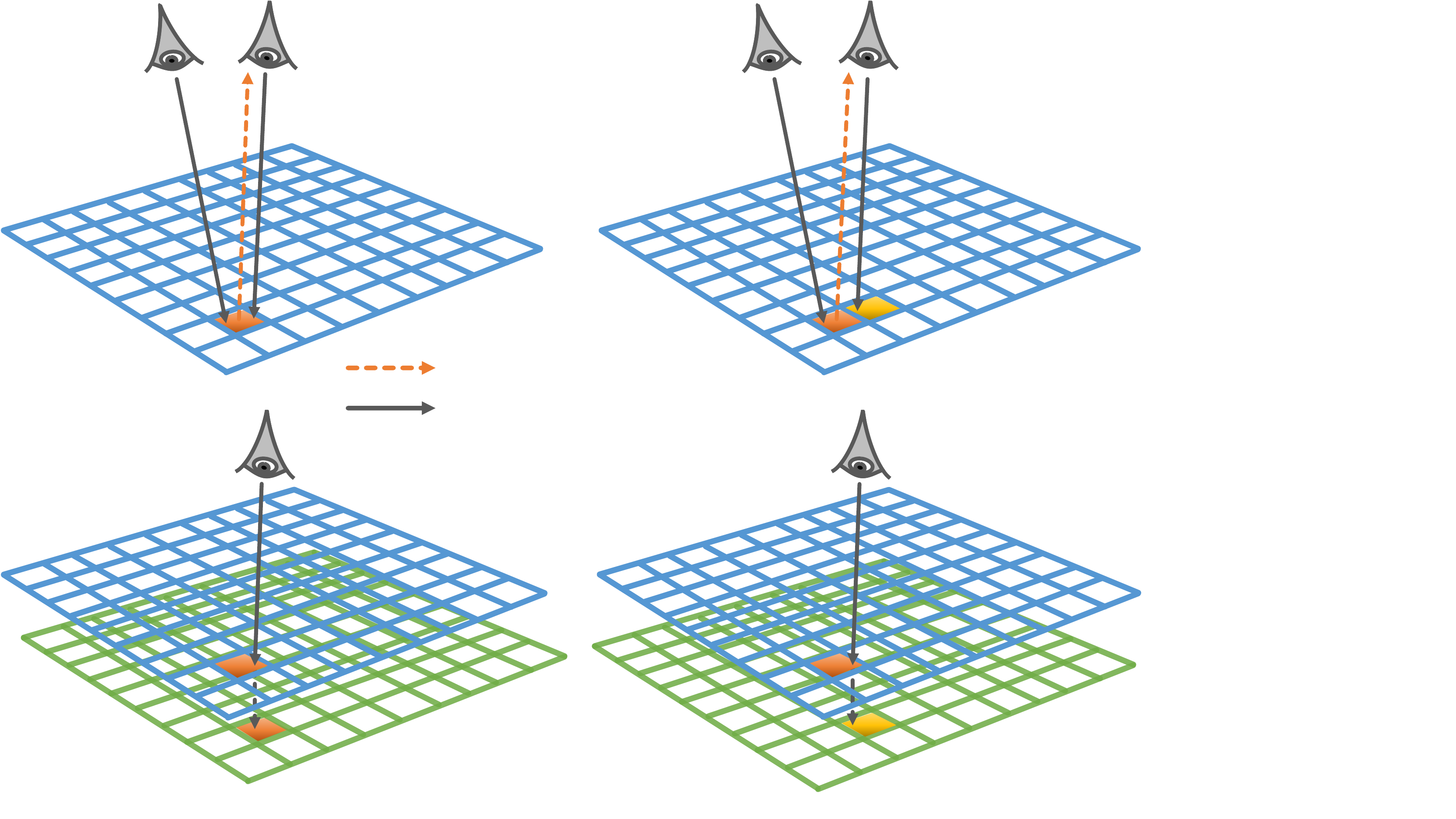}};
    \node[anchor=south west] at (0.0, 5.55) { \textcolor{black}{Frame $t$} };
    \node[anchor=south west] at (2.05, 5.55) { \textcolor{black}{Frame $t-1$} };
    \node[anchor=south west] at (4.25, 5.55) { \textcolor{black}{Frame $t$} };
    \node[anchor=south west] at (6.35, 5.55) { \textcolor{black}{Frame $t-1$} };
    \node[anchor=south west] at (2.8, 4.55) { \textcolor{black}{Shadow map (SM)} };
    \node[anchor=south west] at (0.2, 5.95) { \textcolor{black}{\small A. Small camera movement} };
    \node[anchor=south west] at (4.6, 5.9) { \textcolor{black}{\small B. Large camera movement} };
    \node[anchor=south west] at (3.1, 3.0) { \textcolor{black}{Pixel reprojection} };
    \node[anchor=south west] at (3.1, 2.7) { \textcolor{black}{SM lookup} };
    \node[anchor=south west] at (3.1, 2.0) { \textcolor{black}{Perturbation 0} };
    \node[anchor=south west] at (3.1, 0.4) { \textcolor{black}{Perturbation 1} };
    \node[anchor=south west] at (0.3, -0.3) { \textcolor{black}{\small C. Small SM perturbation} };
    \node[anchor=south west] at (4.9, -0.3) { \textcolor{black}{\small D. Large SM perturbation} };
\end{tikzpicture}
\caption{Comparing the effect of motion-vector and perturbation loss on shadow map texture lookup. A and C indicate that a small camera movement in time is equivalent to a small perturbation of SM in space. Similarly, B and D indicate that a large camera movement is same as large perturbation of SM in space.}
\label{fig:perturbationLossEquivalance}
\vspace{50pt}
\end{figure}

In this section, we compare the similarities an dissimilarities between our perturbation loss with motion vector \cite{NeuralSS} based loss function. A motion-vector finds a pixel's location in the previous frame by projecting the pixel's world-space position on the previous frame. A motion-vector based loss thus computes the motion-vector compensated temporal pixel difference in the network output as error which is then backpropagated through the network for learning. Note that a motion-vector based loss uses historical data \textbf{only} during training.

We argue that our perturbation loss has similar end effect as motion vector based loss while offering more control over temporal stability and simplicity in data collection. In case of motion vector loss, due to discrete nature of pixels, the reprojected pixel in the previous frame may correspond to a slightly different world-space location. Assuming the emitter is fixed, a small difference in world space value of a pixel across time may result in different shadow-map texture lookup, as shown in figure \ref{fig:perturbationLossEquivalance} (b). This is equivalent to perturbing the emitter (our approach) while keeping the camera position fixed as shown in (d). However, for motion-vector loss, if the framerate is too high, the world-space difference in pixel may be small. The pixel and its reprojection may share the same shadow map texel, as shown in (a). Same is true if the shadow map texels are large. In both cases, the temporal error is zero and the network does not learn. Conversely, if the framerate is too low, we may not find a valid reprojection and the error must be forced to zero. As such, a careful balance of the framerate is required for motion-vector losses. However, in our case, we can directly adjust the emitter perturbation such that the differences are large enough for non-zero backpropagation as shown in figure (d). This is achieved by by setting the perturbation magnitude proportional to the distance of emitter from the scene and size of the emitter. Additionally, motion-vector losses rely on long sequence of key-framed images which adds additional complexity to data collection pipeline. Our approach do not rely on temporal information, as such we can sample the scene with arbitrary camera, emitter and object trajectories with any desired framerate.

\section{Performance optimization}

We discuss an interesting aspect of UNet architecture that we did not discuss in the main document.

\begin{table}[t!]
\caption{\label{tab:netUnOpLayerParam} Layer-wise compute, learnable parameters, and temporary storage read/write access for a 5-layer network processing 1024x2048 resolution inputs.}
\begin{tabular}{|c|c|c|c|c|}
\hline
\# &
\begin{tabular}[c]{@{}c@{}}runtime(ms)/\\ resolution\end{tabular} & \begin{tabular}[c]{@{}c@{}}Compute\\ (GFlops)\end{tabular} & \begin{tabular}[c]{@{}c@{}}\# parameters\\ ($\times10^3$)\end{tabular} & \begin{tabular}[c]{@{}c@{}}Temp. storage IO\\  (MPixels)\end{tabular} \\ \hline
0 & \begin{tabular}[c]{@{}c@{}}11.0ms\\ (1Kx2K)\end{tabular} & 7.15 & 3.4 & 278.9 \\ \hline
1 & \begin{tabular}[c]{@{}c@{}}3.67ms\\ (512x1K)\end{tabular} & 8.05 & 15.4 & 117.4 \\ \hline
2 & \begin{tabular}[c]{@{}c@{}}1.63ms\\ (256x512)\end{tabular} & 8.05 & 61.4 & 58.7 \\ \hline
3 & \begin{tabular}[c]{@{}c@{}}0.79ms\\ (128x256)\end{tabular} & 8.05 & 245.8 & 29.3 \\ \hline
4 & \begin{tabular}[c]{@{}c@{}}0.17ms\\ (64x128)\end{tabular} & 2.68 & 327.68 & 6.29 \\ \hline
\end{tabular}
\end{table}

From table \ref{tab:netUnOpLayerParam}, we see that the runtime performance of our network is proportional to the temporary storage, not the number of compute operations. Notice how the compute flops are nearly constant for the first 4 layers, yet the runtime drops as we go down the layers. This indicates the performance is bounded by memory bandwidth. Required memory bandwidth is proportional to the size of temporary buffers and how well the buffers are cached. Caching is most effective when the size of the temporary buffers are small. In fact due to higher cache misses in the layer 0, we see a more than linear growth in runtime compared to layer 1. We verified this using a profiler. A detailed layerwise breakdow of the network runtime (without optimization) is provided in table \ref{tab:netUnOpLayerParam}.

\begin{table}[h]
\caption{\label{tab:netUnOpLayerTiming} Measured layer-wise compute time for a 5-layer network processing 1024x2048 resolution input.}
\begin{tabular}{|c|c|c|c|}
\hline
\begin{tabular}[c]{@{}c@{}}layer/\\ Resolution\end{tabular} & \begin{tabular}[c]{@{}c@{}}Encoder\\ Op/Time(ms)\end{tabular} & \begin{tabular}[c]{@{}c@{}}Decoder\\ Op/Time(ms)\end{tabular} & \begin{tabular}[c]{@{}c@{}}Total\\ (ms)\end{tabular} \\ \hline
\begin{tabular}[c]{@{}c@{}}0\\ (1Kx2K)\end{tabular} & \begin{tabular}[c]{@{}c@{}}Conv2d + DnSamp\\ (4.27 + 0.72)\end{tabular} & \begin{tabular}[c]{@{}c@{}}UpSamp + Skip + Conv2d\\ (0.24 + 0.37 + 5.37)\end{tabular} & 11.0 \\ \hline
\begin{tabular}[c]{@{}c@{}}1\\ (512x1K)\end{tabular} & \begin{tabular}[c]{@{}c@{}}Conv2d + DnSamp\\ (1.38 + 0.63)\end{tabular} & \begin{tabular}[c]{@{}c@{}}UpSamp + Skip + Conv2d\\ (0.12 + 0.19 + 1.36)\end{tabular} & 3.67 \\ \hline
\begin{tabular}[c]{@{}c@{}}2\\ (256x512)\end{tabular} & \begin{tabular}[c]{@{}c@{}}Conv2d + DnSamp\\ (0.58 + 0.33)\end{tabular} & \begin{tabular}[c]{@{}c@{}}UpSamp + Skip + Conv2d\\ (0.07 + 0.10 + 0.55)\end{tabular} & 1.63 \\ \hline
\begin{tabular}[c]{@{}c@{}}3\\ (128x256)\end{tabular} & \begin{tabular}[c]{@{}c@{}}Conv2d + DnSamp\\ (0.29 + 0.10)\end{tabular} & \begin{tabular}[c]{@{}c@{}}UpSamp + Skip + Conv2d\\ (0.04 + 0.05 + 0.31)\end{tabular} & 0.79 \\ \hline
\begin{tabular}[c]{@{}c@{}}4\\ (64x128)\end{tabular} & \multicolumn{2}{c|}{\begin{tabular}[c]{@{}c@{}}Conv2d\\ (0.17)\end{tabular}} & 0.17 \\ \hline
\end{tabular}
\end{table}

Further performance improvements may be possible through pruning \cite{lotteryTicketML} of the network weights. Performance of our network largely depends on the size of temporary buffers; not how the buffers are connected. As such, simply pruning the weights may not improve performance unless we also reduce the size of temporary storage. Another avenue for exploration is lowering the precision of temporary storage to 8 bits as the network output is always between 0 and 1. We leave these optimizations for further exploration as future work.

\section{Network depth optimization}

In this section, we first derive the mathematical formula for penumbra width using our simplified model as described in the main paper. We estimate the parameters $\theta$, and $\theta_\delta$ as follows:

\begin{figure}[b!]
\begin{tikzpicture}
    \node[anchor=south west,inner sep=0] at (0,0){
    \includegraphics[width=11.5cm, trim={0cm 0.5cm 0 0cm},clip]{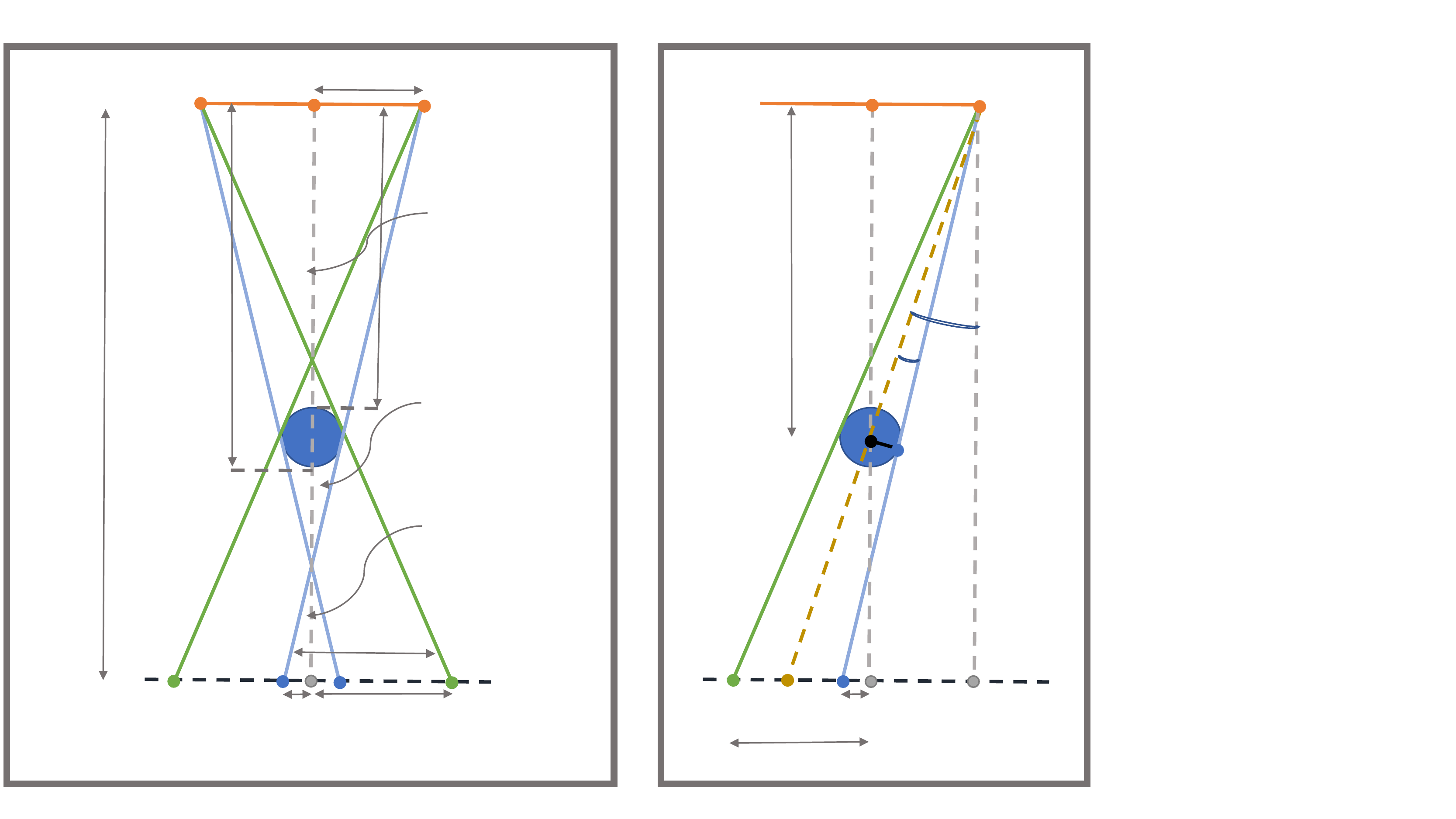}};
    \node[anchor=south west] at (0.75, 5.95) { \textcolor{black}{\small a. Model estimating penumbra}};
    \node[anchor=south west] at (2.75, 5.525) { \textcolor{black}{$r_e$}};
    \node[anchor=south west] at (1.0, 5.45) { \textcolor{black}{Emitter}};
    \node[anchor=south west] at (3.25, 4.45) { \textcolor{black}{Center line}};
    \node[anchor=south west] at (2.95, 3.525) { \textcolor{black}{$z_{min}$}};
    \node[anchor=south west] at (1.05, 3.75) { \textcolor{black}{$z_{max}$}};
    \node[anchor=south west] at (0.4, 2.8) { \textcolor{black}{$z_f$}};
    \node[anchor=south west] at (3.25, 2.9) { \textcolor{black}{$x_a < 0$}};
    \node[anchor=south west] at (3.25, 1.9) { \textcolor{black}{$x_a > 0$}};
    \node[anchor=south west] at (3.4, 1.3) { \textcolor{black}{Penumbra}};
    \node[anchor=south west] at (3.4, 0.95) { \textcolor{black}{$\approx x_a + x_b$}};
    \node[anchor=south west] at (2.1, 0.45) { \textcolor{black}{$x_a$}};
     \node[anchor=south west] at (3.0, 0.40) { \textcolor{black}{$x_b$}};
     
     \node[anchor=south west] at (5.35, 5.95) { \textcolor{black}{\small b. Estimating parameters}};
     \node[anchor=south west] at (6.7, 5.45) { \textcolor{black}{\small P}};
     \node[anchor=south west] at (7.5, 5.45) { \textcolor{black}{\small A}};
     \node[anchor=south west] at (5.7, 4.0) { \textcolor{black}{$z_m$}};
     \node[anchor=south west] at (7.3, 3.35) { \textcolor{black}{$\theta$}};
     \node[anchor=south west] at (6.9, 2.95) { \textcolor{black}{$\theta_{\delta}$}};
     \node[anchor=south west] at (6.5, 2.65) { \textcolor{black}{\small M}};
     \node[anchor=south west] at (7.05, 2.55) { \textcolor{black}{\small D}};
      \node[anchor=south west] at (6.05, 0.5) { \textcolor{black}{\small B}};
     \node[anchor=south west] at (6.9, 0.5) { \textcolor{black}{\small Q}};
     \node[anchor=south west] at (6.55, 0.45) { \textcolor{black}{$x_a$}};
      \node[anchor=south west] at (7.5, 0.5) { \textcolor{black}{\small C}};
      \node[anchor=south west] at (6.05, 0.05) { \textcolor{black}{$x_b$}};
\end{tikzpicture}
\caption{\label{fig:estimatePenubraWidth} A simplified model to estimate the penumbra size at a given pixel. We assume our occluder is spherical in shape, forming a convex bounding-sphere around the occlusion geometry as shown in figure (a) on the left. Figure (b) shows a simplified diagram to estimate the parameters $\theta, \theta_\delta$.}
\end{figure}

\begin{figure*}[t!]
\begin{tikzpicture}
    \node[anchor=south west,inner sep=0] at (0,0){
     \includegraphics[width=18.5cm,trim={1cm 4.0cm 1 0cm},clip]{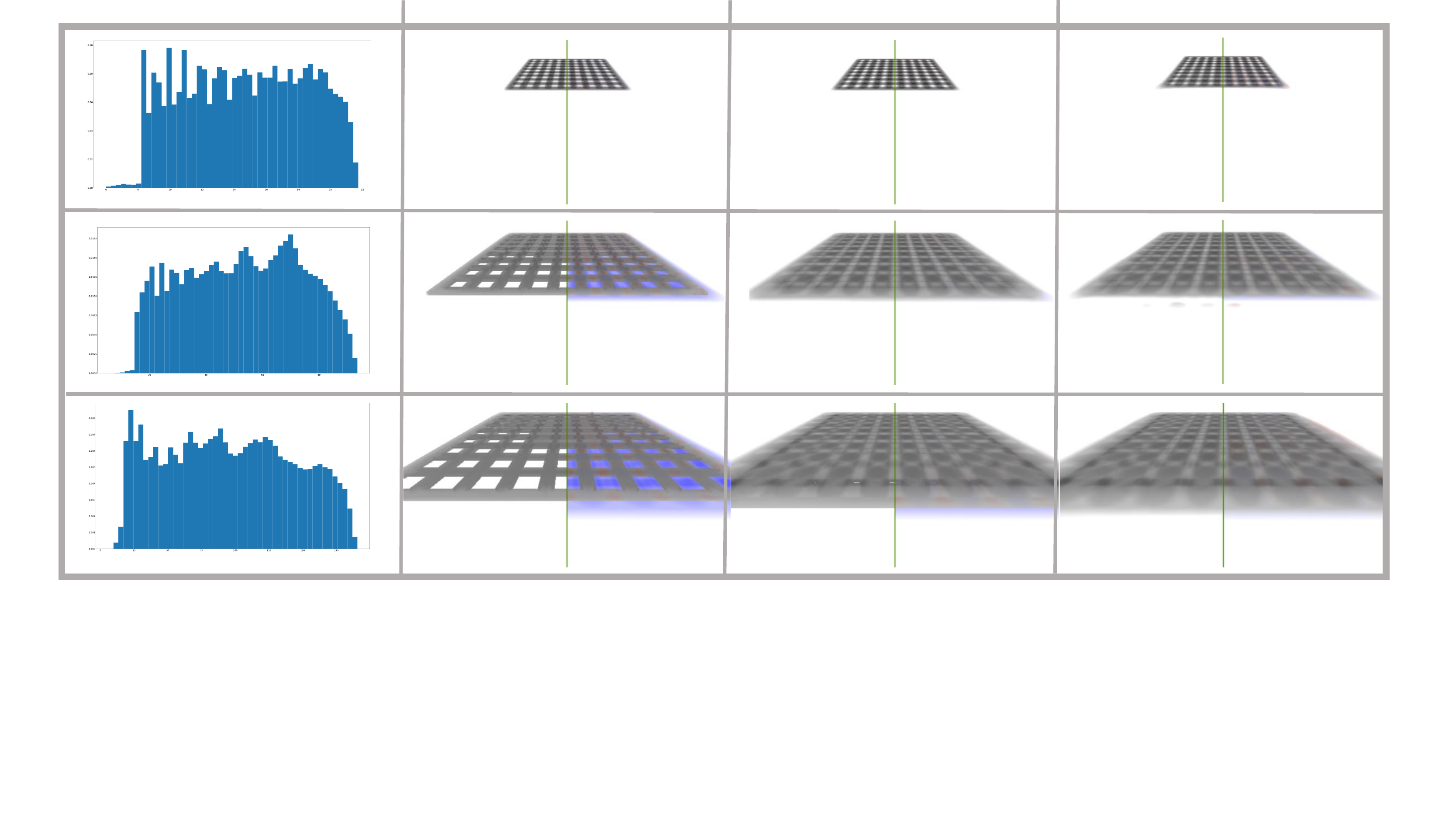}};
     \node[anchor=south west] at (1.55, 8.5) { \textcolor{black}{Predicted}};
     \node[anchor=south west] at (0.4, 8.1) { \textcolor{black}{penumbra width distribution} };
     \node[anchor=south west] at (6, 8.5) { \textcolor{black}{3-layers,}};
      \node[anchor=south west] at (5.1, 8.1) { \textcolor{black}{Receptive field: 24 pixels} };
      \node[anchor=south west] at (10.4, 8.5) { \textcolor{black}{5-layers,}};
      \node[anchor=south west] at (9.5, 8.1) { \textcolor{black}{Receptive field: 96 pixels} };
      \node[anchor=south west] at (14.6, 8.5) { \textcolor{black}{7-layers,}};
      \node[anchor=south west] at (13.7, 8.1) { \textcolor{black}{Receptive field: 384 pixels} };
       \node[anchor=south west, rotate=90] at (0.7, 6.40) { \textcolor{black}{\small Density} };
        \node[anchor=south west, rotate=90] at (4.75, 5.73) { \textcolor{black}{\small $P_{95}$ width: 21 pxls} };
        \node[anchor=south west, rotate=90] at (4.75, 3.33) { \textcolor{black}{\small $P_{95}$ width: 90 pxls} };
        \node[anchor=south west, rotate=90] at (4.75, 0.9) { \textcolor{black}{\small $P_{95}$ width: 180 pxls} };
      \node[anchor=south west, rotate=90] at (0.7, 4.0) { \textcolor{black}{\small Density} };
      \node[anchor=south west, rotate=90] at (0.7, 1.62) { \textcolor{black}{\small Density} };
      \node[anchor=south west] at (0.7, 5.60) { \textcolor{black}{\small Penumbra width (in pixels)} };
      \node[anchor=south west] at (0.7, 3.20) { \textcolor{black}{\small Penumbra width (in pixels)} };
      \node[anchor=south west] at (0.7, 0.82) { \textcolor{black}{\small Penumbra width (in pixels)} };
      \node[anchor=south west] at (7.4, 5.8) { \textcolor{black}{MSE: 0.005} };
      \node[anchor=south west] at (11.75, 5.8) { \textcolor{black}{MSE: 0.005} };
      \node[anchor=south west] at (16.05, 5.8) { \textcolor{black}{MSE: 0.012} };
      \node[anchor=south west] at (7.4, 3.4) { \textcolor{black}{MSE: 0.083} };
      \node[anchor=south west] at (11.75, 3.4) { \textcolor{black}{MSE: 0.009} };
      \node[anchor=south west] at (16.05, 3.4) { \textcolor{black}{MSE: 0.018} };
      \node[anchor=south west] at (7.4, 1.0) { \textcolor{black}{MSE: 0.161} };
      \node[anchor=south west] at (11.75, 1.0) { \textcolor{black}{MSE: 0.042} };
      \node[anchor=south west] at (16.05, 1.0) { \textcolor{black}{MSE: 0.019} };
      \node[anchor=south west] at (0.9, 0.3) { \textcolor{black}{$\mathbf{P_{95}}$ : $95^{th}$ percentile} };
      \node[anchor=south west] at (5.1, 0.3) { \textcolor{black}{\textbf{Left: }Network o/p} };
    \node[anchor=south west] at (7.8, 0.3) { \textcolor{black}{\textbf{Right: }Superimp. error} };
    \node[anchor=south west] at (11.6, 0.36) { \textcolor{red}{Red:} \textcolor{black}{Excess shadow} };
    \node[anchor=south west] at (14.4, 0.29) { \textcolor{blue}{Blue:}\textcolor{black}{Missing shadow} };
\end{tikzpicture}
\caption{Empirical validation of our penumbra width prediction model. We predict the penumbra width on the vertical axis and vary the number of layers on the horizontal axis. The diagonal represents the optimal number of layers and elements below the diagonal show high error as the receptive field is not large enough to accommodate the penumbra width.}
\label{fig:empiricalDepth0}
\end{figure*}

\begin{equation}
    z_m = PM = \frac{z_{max} + z_{min}}{2}
\end{equation}
\begin{equation}
    r_s = MD = \frac{z_{max} - z_{min}}{2}
\end{equation}

From $\triangle AMP$,
\begin{equation}
    AM = \sqrt{z_m^2 + r_e^2}
\end{equation}

From $\triangle AMD$,
\begin{equation}
    \theta_{\delta} = sin^{-1}\frac{MD}{AM} = sin^{-1} \frac{r_s}{\sqrt{z_m^2 + r_e^2}}
\end{equation}

From $\triangle AMP$ and $\triangle BMQ$,
\begin{equation}
    BQ = \frac{AP \cdot MQ}{MP} = \frac{r_e(z_f -z_m)}{z_m}
\end{equation}

From $\triangle ABC$,
\begin{equation}
    \theta = tan^{-1}\frac{BQ + QC}{AC} = tan^{-1}\frac{BQ + r_e}{z_f}
\end{equation}

Therefore,

\begin{equation}
    x_a = z_f\,tan(\theta - \theta_{\delta}) - r_e
\end{equation}
\begin{equation}
    x_b = z_f\,tan(\theta + \theta_{\delta}) - r_e.
\end{equation}

\begin{figure*}[t!]
\begin{tikzpicture}
    \node[anchor=south west,inner sep=0] at (0,0){
     \includegraphics[width=18.5cm,trim={1cm 4.0cm 1 0cm},clip]{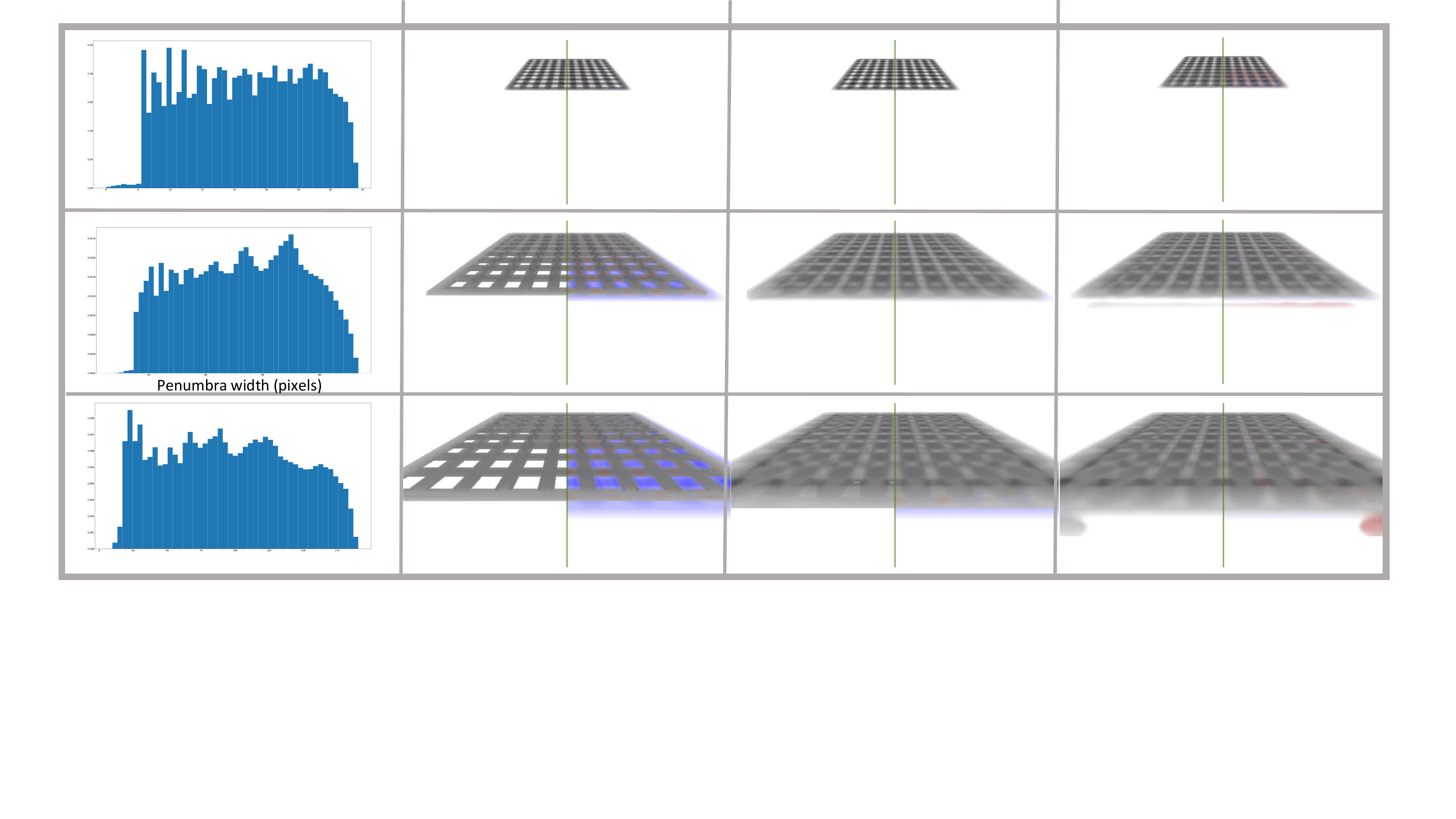}};
     \node[anchor=south west] at (1.55, 8.5) { \textcolor{black}{Predicted}};
     \node[anchor=south west] at (0.4, 8.1) { \textcolor{black}{penumbra width distribution} };
     \node[anchor=south west] at (6, 8.5) { \textcolor{black}{3-layers,}};
      \node[anchor=south west] at (5.1, 8.1) { \textcolor{black}{Receptive field: 24 pixels} };
      \node[anchor=south west] at (10.4, 8.5) { \textcolor{black}{5-layers ,}};
      \node[anchor=south west] at (9.5, 8.1) { \textcolor{black}{Receptive field: 96 pixels} };
      \node[anchor=south west] at (14.6, 8.5) { \textcolor{black}{7-layers,}};
      \node[anchor=south west] at (13.7, 8.1) { \textcolor{black}{Receptive field: 384 pixels} };
       \node[anchor=south west, rotate=90] at (0.7, 6.40) { \textcolor{black}{\small Density} };
        \node[anchor=south west, rotate=90] at (4.75, 5.73) { \textcolor{black}{\small $P_{95}$ width: 21 pxls} };
        \node[anchor=south west, rotate=90] at (4.75, 3.33) { \textcolor{black}{\small $P_{95}$ width: 90 pxls} };
        \node[anchor=south west, rotate=90] at (4.75, 0.9) { \textcolor{black}{\small $P_{95}$ width: 180 pxls} };
      \node[anchor=south west, rotate=90] at (0.7, 4.0) { \textcolor{black}{\small Density} };
      \node[anchor=south west, rotate=90] at (0.7, 1.62) { \textcolor{black}{\small Density} };
      \node[anchor=south west] at (0.7, 5.60) { \textcolor{black}{\small Penumbra width (in pixels)} };
      \node[anchor=south west] at (0.7, 3.20) { \textcolor{black}{\small Penumbra width (in pixels)} };
      \node[anchor=south west] at (0.7, 0.82) { \textcolor{black}{\small Penumbra width (in pixels)} };
      \node[anchor=south west] at (7.4, 5.8) { \textcolor{black}{MSE: 0.005} };
      \node[anchor=south west] at (11.75, 5.8) { \textcolor{black}{MSE: 0.005} };
      \node[anchor=south west] at (16.05, 5.8) { \textcolor{black}{MSE: 0.019} };
      \node[anchor=south west] at (7.4, 3.4) { \textcolor{black}{MSE: 0.083} };
      \node[anchor=south west] at (11.75, 3.4) { \textcolor{black}{MSE: 0.010} };
      \node[anchor=south west] at (16.05, 3.4) { \textcolor{black}{MSE: 0.017} };
      \node[anchor=south west] at (7.4, 1.0) { \textcolor{black}{MSE: 0.161} };
      \node[anchor=south west] at (11.75, 1.0) { \textcolor{black}{MSE: 0.042} };
      \node[anchor=south west] at (16.05, 1.0) { \textcolor{black}{MSE: 0.030} };
      \node[anchor=south west] at (0.9, 0.3) { \textcolor{black}{$\mathbf{P_{95}}$ : $95^{th}$ percentile} };
      \node[anchor=south west] at (5.1, 0.3) { \textcolor{black}{\textbf{Left: }Network o/p} };
    \node[anchor=south west] at (7.8, 0.3) { \textcolor{black}{\textbf{Right: }Superimp. error} };
    \node[anchor=south west] at (11.6, 0.36) { \textcolor{red}{Red:} \textcolor{black}{Excess shadow} };
    \node[anchor=south west] at (14.4, 0.29) { \textcolor{blue}{Blue:}\textcolor{black}{Missing shadow} };
\end{tikzpicture}
\caption{Empirical validation of our penumbra width prediction model. We predict the penumbra width on the vertical axis and vary the number of layers on the horizontal axis while keeping the compute flops constant across networks. The flops constrained networks show similar behavior as the unconstrained networks in figure \ref{fig:empiricalDepth0}.}
\label{fig:empiricalDepth2}
\end{figure*}

\begin{figure*}[t!]
\begin{tikzpicture}
    \node[anchor=south west,inner sep=0] at (0,0){
\includegraphics[width=20.25cm, trim={0cm 0.0cm 0 0cm},clip]{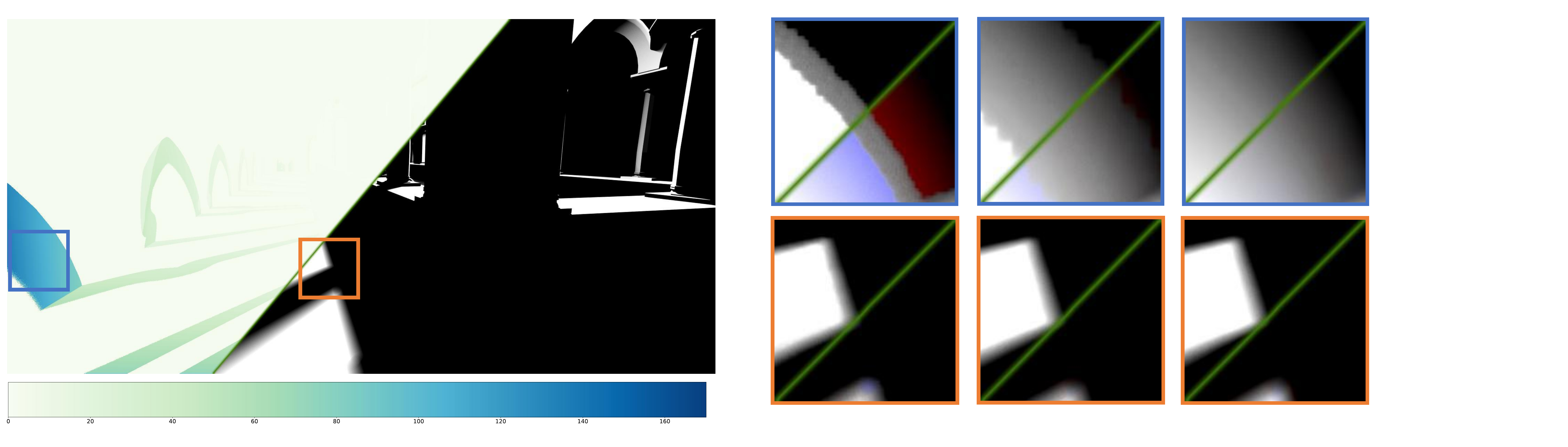}};
\node[anchor=south west] at (0.5, 5.4) { \textcolor{black}{Predicted penumbra width (in screen space)} };
\node[anchor=south west] at (6.5, 5.4) { \textcolor{black}{Ray-traced reference} };
\node[anchor=south west] at (10.45, 5.75) { \textcolor{black}{3-layers}};
\node[anchor=south west] at (10.25, 5.4) { \textcolor{black}{Rf: 24 pixels} };
\node[anchor=south west] at (13.15, 5.75) { \textcolor{black}{5-layers}};
\node[anchor=south west] at (12.95, 5.4) { \textcolor{black}{Rf: 96 pixels} };
\node[anchor=south west] at (15.85, 5.75) { \textcolor{black}{7-layers}};
\node[anchor=south west] at (15.55, 5.4) { \textcolor{black}{Rf: 384 pixels} };
\node[anchor=south west, rotate=90] at (9.70, 1.25) { \textcolor{black}{Predicted penumbra width}};
\node[anchor=south west, rotate=90] at (10.05, 3.55) { \textcolor{black}{130 pixels}};
\node[anchor=south west, rotate=90] at (10.05, 1.15) { \textcolor{black}{23 pixels}};
\node[anchor=south west] at (10.45, 5.1) { \textcolor{white}{MSE: 0.232} };
\node[anchor=south west] at (13.1, 5.1) { \textcolor{white}{MSE: 0.051} };
\node[anchor=south west] at (15.75, 5.1) { \textcolor{white}{MSE: 0.015} };
\node[anchor=south west] at (10.45, 2.5) { \textcolor{white}{MSE: 0.009} };
\node[anchor=south west] at (13.1, 2.5) { \textcolor{white}{MSE: 0.008} };
\node[anchor=south west] at (15.75, 2.5) { \textcolor{white}{MSE: 0.010} };
\node[anchor=south west] at (9.5, 0.05) { \textcolor{black}{\textbf{Left: }Network o/p} };
\node[anchor=south west] at (12.05, 0.00) { \textcolor{black}{\textbf{Right: }Superimp. error} };
\node[anchor=south west] at (15.2, 0.00) { \textcolor{black}{\textbf{Rf: }Receptive field} };
\node[anchor=south west] at (11.0, -0.35) { \textcolor{red}{Red:} \textcolor{black}{Excess shadow} };
    \node[anchor=south west] at (14, -0.42) { \textcolor{blue}{Blue:}\textcolor{black}{Missing shadow} };
    \node[anchor=south west] at (3.0, -0.25) { \textcolor{black}{Penumbra width (in pixels)} };
\end{tikzpicture}
\caption{Figure showing the accuracy of our penumbra width prediction model on Sponza. Region with high penumbra width (blue cutout), as predicted by our model is reproduced accurately only by the 7-layer network.}
\label{fig:empiricalDepth1}
\end{figure*}

\subsection{Empirical verification}

\begin{table}[h]
\caption{\label{tab:networkCost0}  Table showing the compute cost and the number of learnable parameters for 3,5,and 7 layer network in figure \ref{fig:empiricalDepth0}.}
\begin{tabular}{|c|c|c|}
\hline
Network & Flops per pixel & \# learnable parameters \\ \hline
3-layer & 8528 & 39.25K \\ \hline
5-layer & 16208 & 653.7K \\ \hline
7-layer & 21968 & 9.501M \\ \hline
\end{tabular}
\end{table}

In figure \ref{fig:empiricalDepth0}, we move a mesh object between a ground plane and an emitter producing penumbra of varying size. For each row in the figure, we predict the distribution of penumbra size and find the maximum (95th percentile) width from the distribution. Across the columns, we vary the receptive field of the network by adjusting the number of layers. Collecting the errors across all combinations, we notice that the elements below the diagonal have high error as the receptive field of the network is not large enough to accommodate the penumbra size. Note that in figure \ref{fig:empiricalDepth0}, networks do not have the same compute flops per pixel as shown in table \ref{tab:networkCost0}. 

\begin{table}[b!]
\caption{\label{tab:networkCost1}  Table showing the compute cost and the number of learnable parameters for 3,5,and 7 layer network shown in figure \ref{fig:empiricalDepth2}.}
\begin{tabular}{|c|c|c|}
\hline
Network & Flops per pixel & \# learnable parameters \\ \hline
3-layer & 14928 & 110.9K \\ \hline
5-layer & 16208 & 653.7K \\ \hline
7-layer & 15328 & 1.626M \\ \hline
\end{tabular}
\vspace{20pt}
\end{table}

Simply increasing the number of layers in a network also increases
the compute cost. As such, any reduction in error might be attributed
to the increased compute requirement. We verify that the reduced
error is indeed due to increased receptive field of the network and
not necessarily due to the increased computation. As such, we constrain the network flops across the 3, 5, and 7 layer network as shown in table \ref{tab:networkCost1}. From figure \ref{fig:empiricalDepth2}, we see that the error reduces along the diagonal despite using networks that are computationally equal (almost).

In figure \ref{fig:empiricalDepth1}, we take a single frame from the \textsc{Sponza} scene and analyze the validity of our penumbra width prediction across different regions in the frame. The increasing accuracy of our network output with the number of layers in a region with large penumbra (yellow cutout) is a strong validation in support of our model.

 \begin{figure*}[t!]
\begin{tikzpicture}
    \node[anchor=south west,inner sep=0] at (0,0){
    \includegraphics[width=\textwidth,trim={0cm 0.9cm 0 0cm},clip]{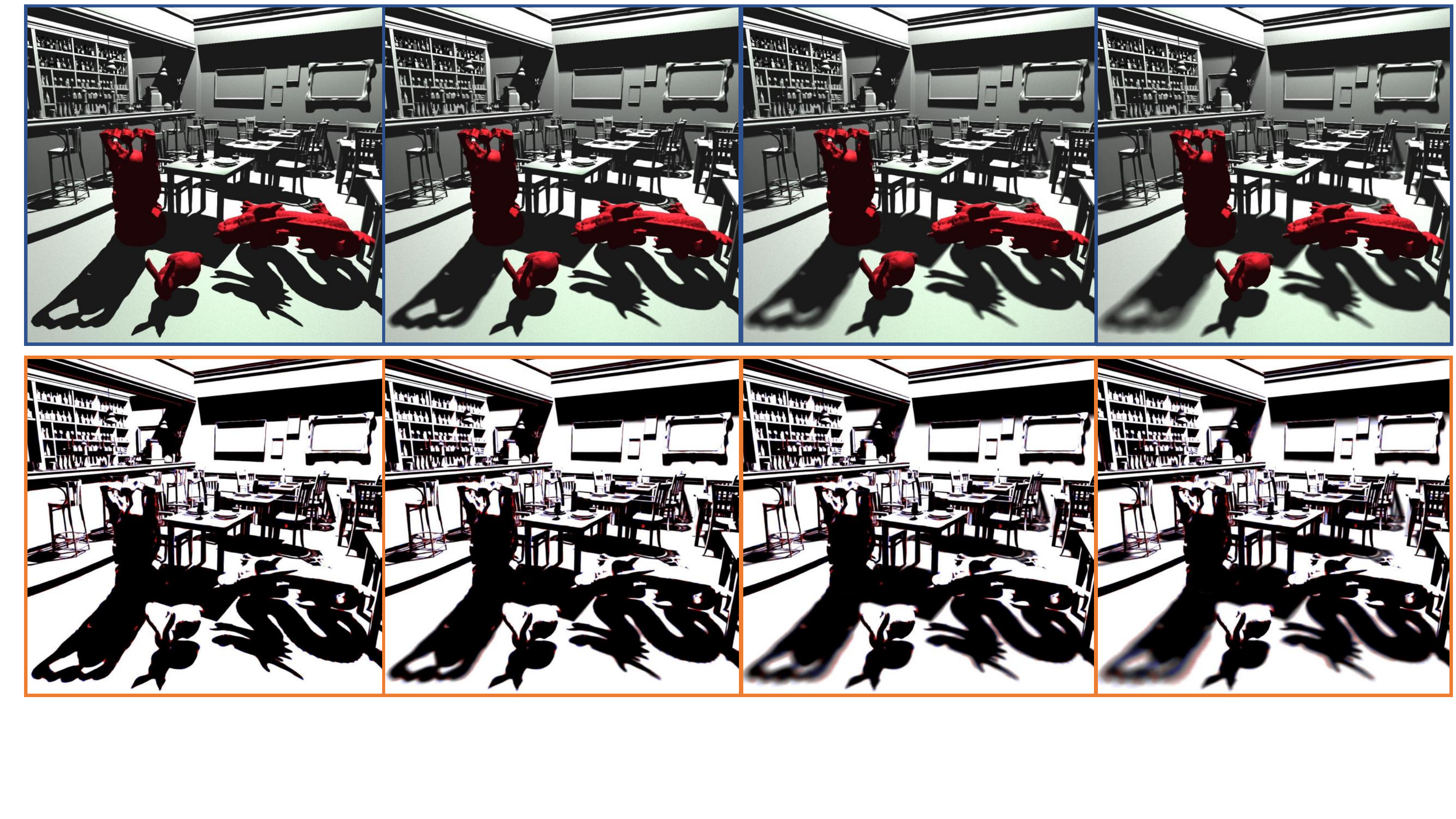}};
    \node[anchor=south west] at (2.4, 9.6) { \textcolor{black}{\huge 0} };
    \node[anchor=south west] at (6.5, 9.6) { \textcolor{black}{\huge 1} };
    \node[anchor=south west] at (11, 9.6) { \textcolor{black}{\huge 2} };
    \node[anchor=south west] at (15.5, 9.6) {\textcolor{black}{\huge 3} };
    \node[anchor=south west, rotate=90] at (0.3, 5.2) {\textcolor{black}{Network o/p (Untrained} \textcolor{red}{objects}\textcolor{black}{)} };
    \node[anchor=south west, rotate=90] at (0.3, 1.8) {\textcolor{black}{Superimposed error} };
    \node[anchor=south west] at (2.0, 1.05) { \textcolor{black}{MSE: 0.095} };
    \node[anchor=south west] at (6.4, 1.05) { \textcolor{black}{MSE: 0.096} };
    \node[anchor=south west] at (10.75, 1.05) { \textcolor{black}{MSE: 0.096} };
    \node[anchor=south west] at (15.1, 1.05) {\textcolor{black}{MSE: 0.097} };
    
    \node[anchor=south west] at (6, 0.45) { \textcolor{red}{Red:} \textcolor{black}{Excess shadow} };
    \node[anchor=south west] at (9.2, 0.38) { \textcolor{blue}{Blue:} \textcolor{black}{Missing shadow} };
\end{tikzpicture}
\vspace{-25pt}
    \caption{Demonstration of the effectiveness of our technique on untrained objects marked in \textcolor{red}{red} in the first row. Second row shows the comparison w.r.t reference with error superimposed over raw network output. Going from left to right, hard-shadow is indexed with 0 while 3 indicates an emitter of diameter 50cm. Shading is applied post-process in the first row.}
\label{fig:resultsUntrainedObjects}
\end{figure*}

\section{Data generation, training and inference}

\begin{figure*}[t!]
\begin{tikzpicture}
    \node[anchor=south west,inner sep=0] at (0,-0.25){
    \includegraphics[width=\textwidth,trim={0cm 10.9cm 0 0cm},clip]{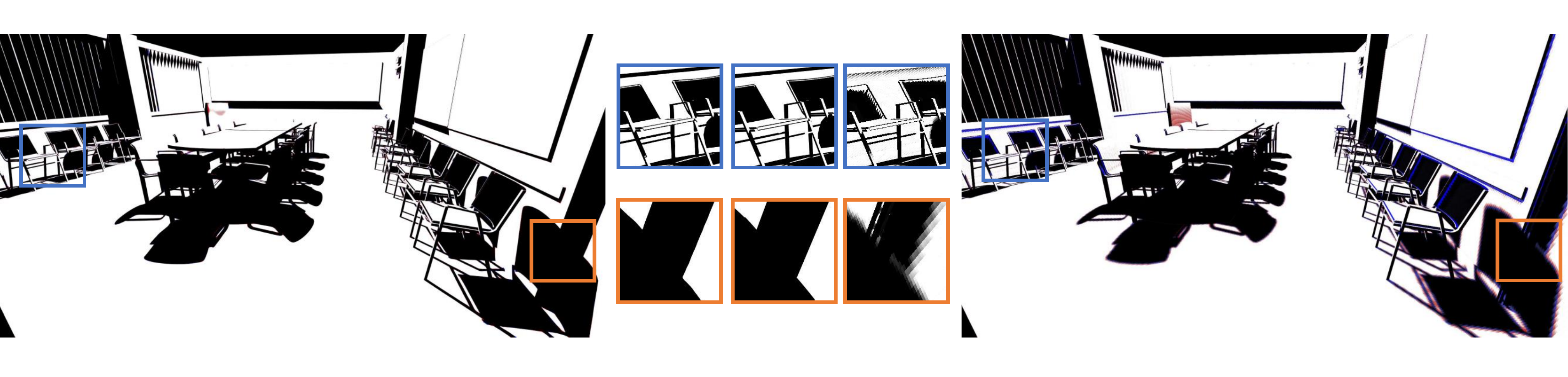}};
    \node[anchor=south west] at (1.8, 3.6) {\textcolor{black}{Ours, 8ms, MSE: 0.035} };
    \node[anchor=south west] at (7.5+0.5, 3.75) {\textcolor{black}{Hard-shadow} };
    \node[anchor=south west] at (13, 3.6) {\textcolor{black}{MSM-3, 4ms, MSE: 0.105} };
    \node[anchor=south west] at (6.9, 3.3) {\textcolor{black}{MSE: 0.055} };
    \node[anchor=south west] at (9.4, 3.3) {\textcolor{black}{MSE: 0.219} };
    \node[anchor=south west] at (7.2, 1.75) {\textcolor{black}{Ours} };
    \node[anchor=south west] at (8.2, 1.75) {\textcolor{black}{Reference} };
    \node[anchor=south west] at (9.7, 1.75) {\textcolor{black}{MSM-3} };
    \node[anchor=south west] at (6.9, 0.2) {\textcolor{black}{MSE: 0.030} };
    \node[anchor=south west] at (9.4, 0.2) {\textcolor{black}{MSE: 0.133} };
    
    \node[anchor=south west,inner sep=0] at (0,-0.45-8.0){
    \includegraphics[width=\textwidth,trim={0cm 3.0cm 0 0cm},clip]{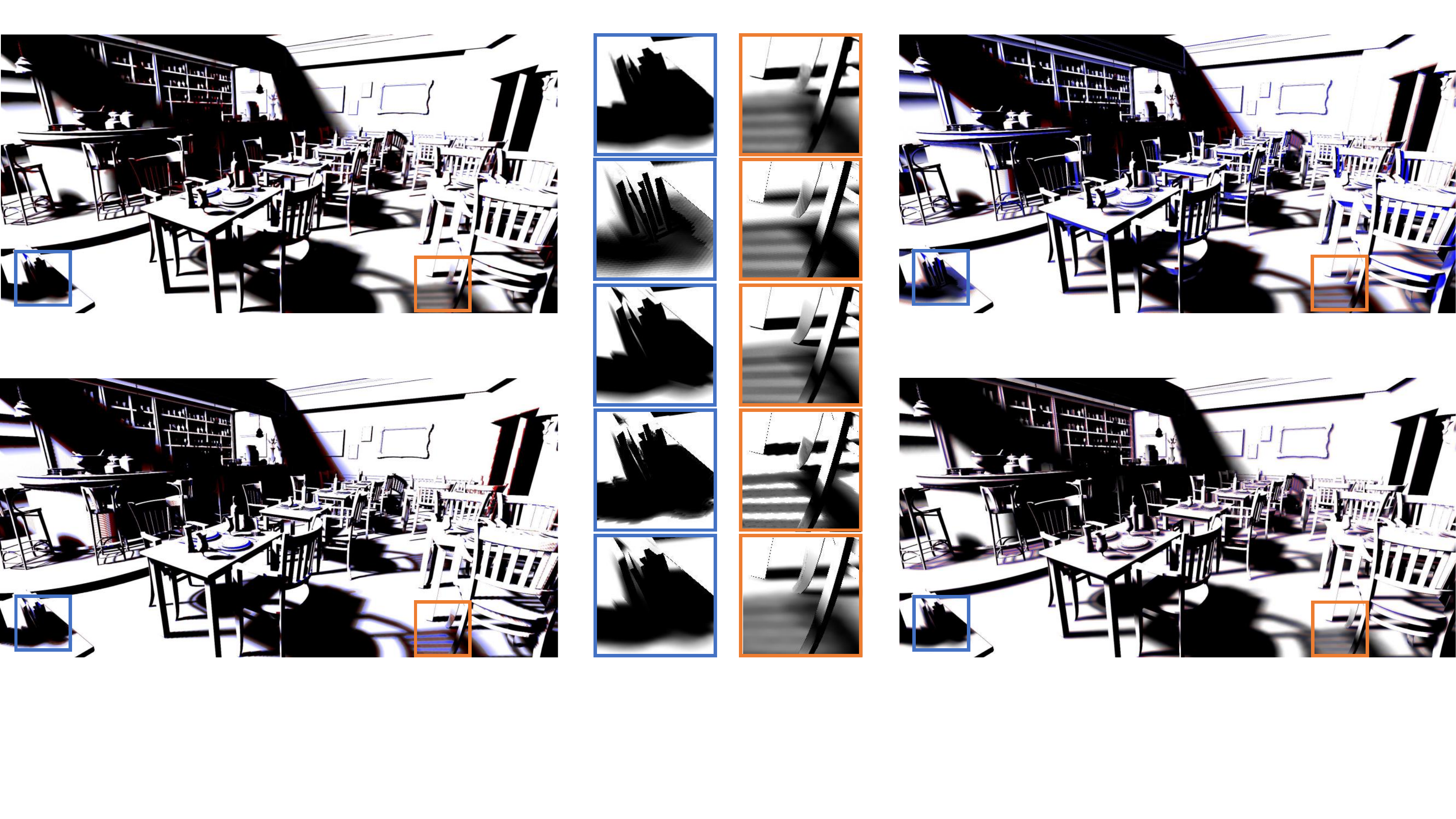}};
    \node[anchor=south west] at (1.8-0.05, 7.5-8.0) {\textcolor{black}{Ours, 8ms, MSE: 0.083} };
    \node[anchor=south west] at (7.5-0.05+0.5, 7.65-8.0) {\textcolor{black}{Soft-shadow} };
    \node[anchor=south west] at (13-0.05, 7.5-8.0) {\textcolor{black}{MSM-9, 4.5ms, MSE: 0.161} };
    \node[anchor=south west] at (1.8-0.05, 3.35-8.0) {\textcolor{black}{PCSS, 5.5ms, MSE: 0.111} };
     \node[anchor=south west] at (12.5-0.05, 3.35-8.0) {\textcolor{black}{5SPP + SVGF, 11.5ms, MSE: 0.094} };
    \node[anchor=south west, rotate=90] at (9.17-0.05, 6.4-8.0) {\textcolor{black}{Ours} };
    \node[anchor=south west, rotate=90] at (9.17-0.05, 4.8-8.0) {\textcolor{black}{MSM-9} };
    \node[anchor=south west, rotate=90] at (9.17-0.05, 3-8.0) {\textcolor{black}{Reference} };
     \node[anchor=south west, rotate=90] at (9.17-0.05, 1.8-8.0) {\textcolor{black}{PCSS} };
     \node[anchor=south west, rotate=90] at (9.17-0.05, -0.15-8.0) {\textcolor{black}{5 SPP+SVGF} };
     \node[anchor=south west, rotate=90] at (7.35-0.05, 6.1-8.0) {\textcolor{black}{MSE: 0.085} };
    \node[anchor=south west, rotate=90] at (7.35-0.05, 4.6-8.0) {\textcolor{black}{MSE: 0.187} };
    \node[anchor=south west, rotate=90] at (7.35-0.05, 1.5-8.0) {\textcolor{black}{MSE: 0.113} };
     \node[anchor=south west, rotate=90] at (7.35-0.05, -0.1-8.0) {\textcolor{black}{MSE: 0.093} };
     \node[anchor=south west, rotate=90] at (11.0-0.05, 6.1-8.0) {\textcolor{black}{MSE: 0.065} };
    \node[anchor=south west, rotate=90] at (11.0-0.05, 4.6-8.0) {\textcolor{black}{MSE: 0.159} };
    \node[anchor=south west, rotate=90] at (11.0-0.05, 1.5-8.0) {\textcolor{black}{MSE: 0.217} };
     \node[anchor=south west, rotate=90] at (11.0-0.05, -0.1-8.0) {\textcolor{black}{MSE: 0.075} };
     
     \node[anchor=south west] at (4.15, -0.63-8.0) { \textcolor{black}{\textbf{Left/Right: }Superimposed error} };
    \node[anchor=south west] at (9.0, -0.63-8.0) { \textcolor{black}{\textbf{Middle (zoom-ins) : }Network output} };
    \node[anchor=south west] at (5.8, -1.01-8.0) { \textcolor{red}{Red:} \textcolor{black}{Excess shadow} };
    \node[anchor=south west] at (9.0, -1.08-8.0) { \textcolor{blue}{Blue:} \textcolor{black}{Missing shadow} };
     
\end{tikzpicture}
\vspace{-18pt}
\caption{\label{fig:comparisonUntrainedTrajectories} Our network's ability to generalize to unseen trajectories (within the same scene) compared to output from competing techniques (MSM, PCSS, Raytracing+Denoising) for hard and soft shadows. MSM-3 and MSM-9 are Moment Shadow Map variants using 3$\times$3 and 9$\times$9 prefiltering kernels. Conference model \copyright  Anat Grynberg and Greg Ward.}
\vspace{-10pt}
\end{figure*}

Orchestrating scene authoring, data collection, training, inference, and generating the final results is one of the key challenges to this project. We author the scenes using Blender \cite{CiteBlender}, ensuring all keyframed camera trajectories do not go outside the emitter frustum. We keyframe several camera and emitter trajectories covering different scenarios and bake them into the scenes. We then upload the scenes to a Linux cluster to generate our training data. A modified version Falcor \cite{CiteFalcor} based on Vulkan (originally DX12) is used along with Python scripts to collect, process and organize the data for training. We use Pytorch \cite{citePytorch} scripts for training which scans through the data in a random order every epoch. While training, we save \textit{N}($=3$) best models based on a test error. The test consist of a small number (10-15) of handpicked images. We run the test with a probability of 0.01 at each training iteration. Thus the test is run roughly every 100 training iterations. Once the training is complete, we select the best model out of \textit{N} by comparing the error on full training dataset. We also collect several statistics during the training for analysis. During inference, we run our model through new trajectories which may have some overlap with training dataset but are not same. We test the performance of our network on a local machine (AMD 5600X, Nvidia 2080Ti) using Falcor and a Cudnn based solution.

The training data is generated using concurrent shadow mapping and ray-tracing passes. The shadow mapping pass outputs three perturbations of the buffers, obtained by jittering the emitter and camera positions. The ray-tracing pass uses 1500 rays per pixel distributed across 8 or more accumulation passes enabling 8x or more MSAA. The ray-tracing pass simultaneously outputs shadows with 4 different levels of softness. To summarize, each frame consist of 3 perturbations of our input buffers and 4 antialiased ray-traced images with increasing softness.

A visual inspection of the generated data is crucial. We compare the unprocessed rasterized shadows with the ray-traced shadows. A match between rasterized and ray-traced output is desired and the two should have minimal and consistent bias, if any. For example, there may be a difference in the position of shadows between rasterization and ray-tracing due to the offset used for preventing self-intersection. We find rescaling the scenes to the same dimension useful for minimizing bias between rasterization and ray-tracing and generating consistent data. Bias also depend on how we generate the rays - emitter to scene or vice versa. The two may be different depending on how backface culling is configured. We prefer emitter to scene ray-tracing as the setup is closer to shadow-mapping.

\section{Network architecture ablation study}

We discuss the limitations of various other network architecture we implemented. The first variation is inspired by Exponential Shadow Maps \cite{ESM08}, where we learn the parameter $\alpha$ in the depth test approximation $e^{\alpha (z-z_f)}$. To remove non-linearity in the final network layer, we perform the Adam-SGD optimization in log-space and exponentiate the output during inference. However, injecting unprocessed input $z$ directly to the output causes severe shadow-aliasing and our loss function (VGG-19) is ineffective in correcting the aliasing in log-space.  In another variation, we tried denoising the rasterized (from emitter) depth - $z$ using a separate network with ray-traced depths as targets. We split the network in two halves - first half for denosing the depth and second half for processing the output of the first half into final shadows. We trained the two network end to end. However, the architecture failed to generate good quality final output compared to our vanilla network. We also tested other variants of the same idea but were equally ineffective.

\section{Results and comparisons}
\begin{figure*}[t!]
\begin{tikzpicture}
    \node[anchor=south west,inner sep=0] at (0,0){
    \includegraphics[width=\textwidth,trim={0cm 7.3cm 0 0cm},clip]{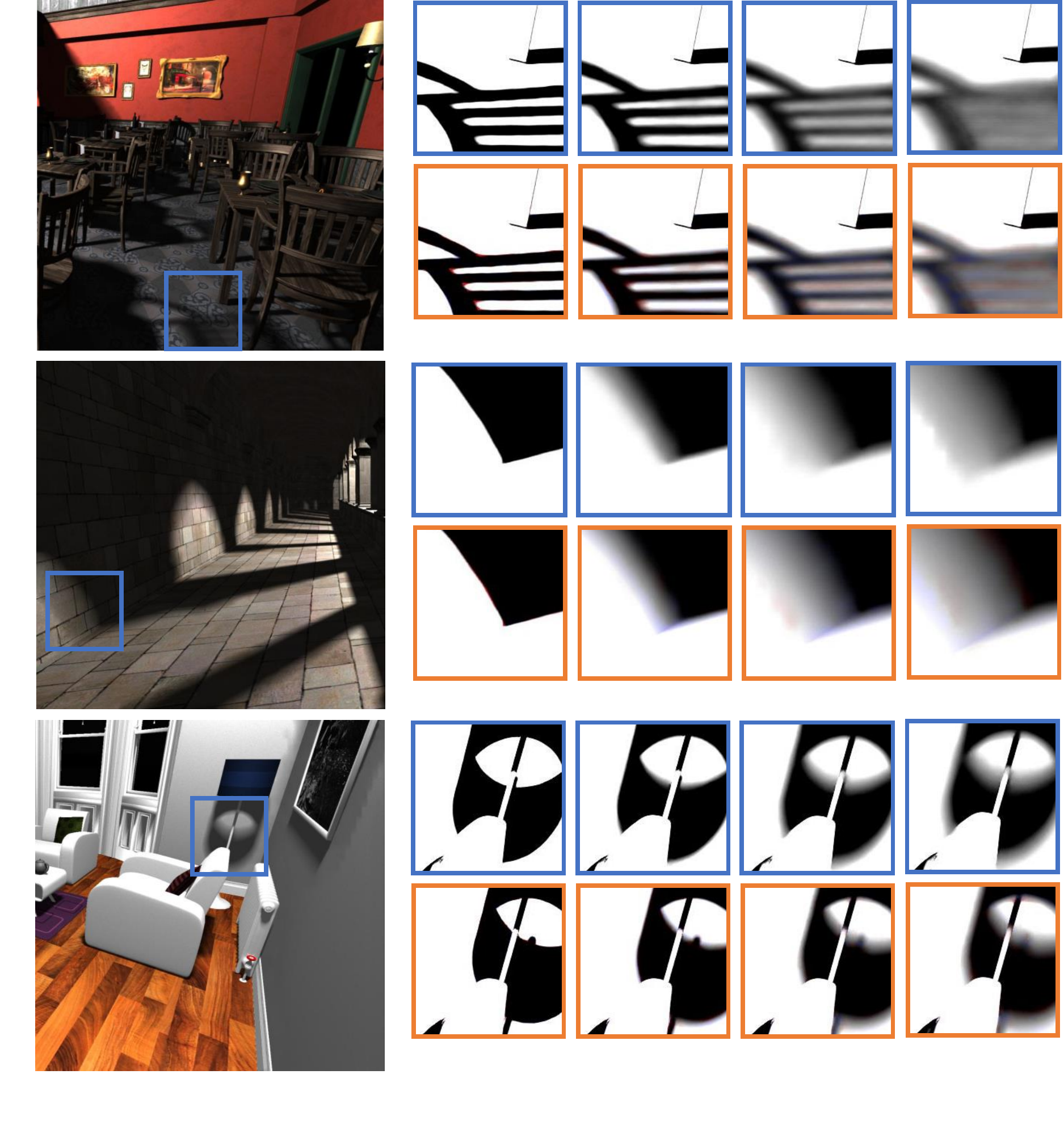}};
    \node[anchor=south west] at (3.5, 18.9) { \textcolor{black}{\huge 4} };
    \node[anchor=south west] at (7.9, 18.9) { \textcolor{black}{\huge 0} };
    \node[anchor=south west] at (10.7, 18.9) { \textcolor{black}{\huge 1} };
    \node[anchor=south west] at (13.4, 18.9) {\textcolor{black}{\huge 2} };
    \node[anchor=south west] at (16.2, 18.9) {\textcolor{black}{\huge 3} };
    
    \node[anchor=south west] at (7.5, 13.1) { \textcolor{black}{MSE: 0.078} };
    \node[anchor=south west] at (10.2, 13.1) { \textcolor{black}{MSE: 0.072} };
    \node[anchor=south west] at (13, 13.1) {\textcolor{black}{MSE: 0.069} };
    \node[anchor=south west] at (15.8, 13.1) {\textcolor{black}{MSE: 0.069} };
    \node[anchor=south west] at (0.8, 13.2) {\textcolor{white}{MSE: 0.070} };
    \node[anchor=south west, rotate=90] at (6.95, 16.65) {\textcolor{black}{Network o/p} };
    \node[anchor=south west, rotate=90] at (6.95, 13.4) {\textcolor{black}{Superimposed error} };
    \node[anchor=south west, rotate=90] at (0.4, 15.0) {\textcolor{black}{ \huge Bistro} };
    
    \node[anchor=south west] at (7.5, 7.1) { \textcolor{black}{MSE: 0.037} };
    \node[anchor=south west] at (10.2, 7.1) { \textcolor{black}{MSE: 0.035} };
    \node[anchor=south west] at (13, 7.1) {\textcolor{black}{MSE: 0.032} };
    \node[anchor=south west] at (15.8, 7.1) {\textcolor{black}{MSE: 0.032} };
    \node[anchor=south west] at (0.8, 12.15) {\textcolor{white}{MSE: 0.034} };
    \node[anchor=south west, rotate=90] at (6.95, 10.65) {\textcolor{black}{Network o/p} };
    \node[anchor=south west, rotate=90] at (6.95, 7.4) {\textcolor{black}{Superimposed error} };
    \node[anchor=south west, rotate=90] at (0.4, 9.5) {\textcolor{black}{ \huge Sponza} };
    
    \node[anchor=south west] at (7.5, 1.1) { \textcolor{black}{MSE: 0.020} };
    \node[anchor=south west] at (10.2, 1.1) { \textcolor{black}{MSE: 0.024} };
    \node[anchor=south west] at (13, 1.1) {\textcolor{black}{MSE: 0.018} };
    \node[anchor=south west] at (15.8, 1.1) {\textcolor{black}{MSE: 0.019} };
    \node[anchor=south west] at (4.6, 1.1) {\textcolor{white}{MSE: 0.019} };
    \node[anchor=south west, rotate=90] at (6.95, 4.65) {\textcolor{black}{Network o/p} };
    \node[anchor=south west, rotate=90] at (6.95, 1.4) {\textcolor{black}{Superimposed error} };
    \node[anchor=south west, rotate=90] at (0.4, 2.5) {\textcolor{black}{ \huge Living room} };
    
    \node[anchor=south west] at (9.5, 0.55) { \textcolor{red}{Red:} \textcolor{black}{Excess shadow} };
    \node[anchor=south west] at (12.7, 0.48) { \textcolor{blue}{Blue:} \textcolor{black}{Missing shadow} };
\end{tikzpicture}
\vspace{-30pt}
    \caption{Figure shows the output of our network (shaded post-process) with varying penumbra sizes in the cutouts (unshaded). Each scene is trained independently and tested on a validation set. The second row in each scene shows the error (superimposed on network oputput) w.r.t. reference. Hard-shadow is indexed 0 while 4 indicates an emitter of diameter 50cm.}
\label{fig:resultsValidation}
\end{figure*}

Figure \ref{fig:comparisonUntrainedTrajectories} compares our technique with other competing techniques on trajectories that were not present in the training.
Figure \ref{fig:resultsValidation} shows the application of our technique across untrained camera and emitter trajectories. For each scene, we train a separate network using a variety of emitter and camera configurations across a range of softness. As such, a single network can generate both hard and soft shadows where the softness is controlled using a scalar input. Figure \ref{fig:resultsUntrainedObjects} shows that our network generalizes across variety of shapes that were not present in the training set.

\section{Limitations and future work}

One of the current limitations of our technique is that it does not naturally extend beyond a single light source. Resolving this is an exciting avenue for future work. Another limitation is it does not generalize well across a \textit{mixture of scenes} with widely different emitter depth distributions. This is primarily due to the (purposefully) compact size of our networks and their limited capacity to fit such multi-modal data. To avoid this, one could potentially train multiple networks across different depth variations and swap between them networks (i.e., based on emitter distance) at runtime. Artifacts also arise when the penumbra sizes exceed the receptive field of the neural network and training tile size; our conservative penumbra size estimate affords us the opportunity, however, to tailor our architecture’s receptive field accordingly. Finally, our perturbation loss can sometimes overblur fine geometric details, such as from thin features like wires, however the gains in temporal stability are significant; fine tuning here can help to mitigate such overblurring.

An exciting avenue for future work involves extending our technique to a finite number of light sources. We can draw inspiration from graph-coloring approaches~\cite{PSSSSGpuPro6} here, where non-overlapping emitter frustums are grouped together in layers and each layer is filtered independently. A neural approach could use a network that takes as input buffers from $n$ sources and outputs the result. During training, we can disable (i.e., zero input) a fraction of the light sources \textit{p} ($< n$) to permit the network to more effectively learn the response from each source, i.e.,  without confounding the effects \textit{between} sources.
 
Finally, runtime optimization of our network is another avenue for exploration. The performance of a UNet depends primarily on memory bandwidth, as opposed to compute operations. A customized pruning technique focused on improving the compute density by reducing the number of temporary buffers is an interesting direction of research. Also the network output is non-HDR, which can be exploited to reduce the size of buffers.

\end{document}